\documentclass[11pt]{article}
\usepackage{jheppub}
\usepackage[utf8]{inputenc}
\usepackage{amsmath}
\usepackage{tikz}
\usepackage{tkz-euclide}
\usepackage{dsfont}
\usepackage{amssymb}
\usepackage{amsmath}
\usepackage{mathrsfs}
\usepackage{appendix}
\usepackage{bbold}
\usepackage{color}

\newcommand{\cA}{\mathcal{A}}
\newcommand{\C}{\mathbb{C}}

\def\ie{\begin{equation}\begin{aligned}}
\def\fe{\end{aligned}\end{equation}}
\begin{document}

\title{5d-4d Correspondence in Twisted M-theory on a Conifold}

\author[a]{Meer Ashwinkumar,}
\author[b,c,d]{Mir Faizal,}
\author[c]{Arshid Shabir,}
\author[d]{Douglas J. Smith,}
\author[e]{Yehao Zhou}
\affiliation[a]{Albert Einstein Center for Fundamental Physics, Institute for Theoretical Physics,
University of Bern, Sidlerstrasse 5, CH-3012 Bern, Switzerland}
\affiliation[b]{Irving K. Barber School of Arts and Sciences,
University of British Columbia - Okanagan, Kelowna,
British Columbia V1V 1V7, Canada}
\affiliation[c]{Canadian Quantum Research Center, 204-3002 32 Ave Vernon, BC V1T 2L7 Canada}
\affiliation[d]{Department of Mathematical Sciences, Durham University,
Upper Mountjoy, Stockton Road, Durham DH1 3LE, UK}
\affiliation[e]{Kavli Institute for the Physics and Mathematics of the Universe (WPI),
University of Tokyo, Kashiwa, Chiba 277-8583, Japan}

\abstract{
We study twisted M-theory in a general conifold background, and describe it in terms of a 5d non-commutative Chern-Simons-matter theory, which is equivalent to 5d non-commutative Chern-Simons theory for a supergroup. In an equivalent description as twisted type IIA string theory, the matter degrees of freedom arise from topological strings stretched between stacks of D6-branes.
In order to study  5d Chern-Simons-matter theories with a boundary, we first construct and investigate the properties of a 4d non-commutative gauged chiral WZW model. We prove the gauge invariant coupling of this 4d theory to the bulk 5d Chern-Simons theory defined on $\mathbb{R}_+ \times \mathbb{C}^2  $, and further generalize our results to the 5d Chern-Simons-matter theory.  We also investigate the toroidal current algebra of the 4d chiral WZW model that arises from radial quantization along one of the complex planes.
Finally, we show that a gauged non-commutative chiral 4d WZW model 
arises from the partition function for quantum 5d non-commutative Chern-Simons theory with boundaries in the BV-BFV formalism, and further generalize this 5d-4d correspondence to the 5d non-commutative Chern-Simons-matter theory for the case of adjoint matter. 
}

\maketitle

\section{Introduction}

Twisted M-theory was first studied by Costello in \cite{Costello:2016nkh}, where it was shown that for a certain twist of M-theory on an 11-dimensional manifold that includes a Taub-NUT background, as well as a particular nontrivial 3-form $C$-field, the physical degrees of freedom can be captured by a 5d non-commutative topological-holomorphic Chern-Simons theory defined on $\mathbb{R} \times \mathbb{C}^2 $. This quantum field theory is holomorphic along $\mathbb{C}^2 $ and topological
along the remaining direction, $\mathbb{R}$.  Moreover, M2-branes in this theory can be described using Wilson lines along $\mathbb{R}$ in this theory, while M5-branes can described using holomorphic surface defects along one of the complex planes. Notably, the Wilson lines can be shown to be associated with representations of variants of a quantum group known as the affine Yangian.

 Twisted M-theory is insensitive to the radius of the 11th circle of M-theory, implying that in the aforementioned background considered by Costello in \cite{Costello:2016nkh}, the theory is equivalent to type IIA string theory with a stack of D6-branes in a $B$-field background, whose world-volume theory is $\Omega$-deformed along two dimensions. The $\Omega$-deformation results in the localization of the D6-brane world-volume theory to the 5d non-commutative Chern-Simons theory. This is in analogy to the localization of the $\Omega$-deformed world-volume theory of D4-branes and D5-branes to (analytically-continued) 3d  and 4d Chern-Simons theory, respectively
\cite{Luo:2014sva,Costello:2018txb}.

In this work, one of our primary aims is to generalize this description of twisted M-theory on a Taub-NUT space to a more general background geometry. In particular, we will initiate the study of twisted M-theory on a general conifold, and propose a description in terms of an extension of 5d non-commutative Chern-Simons theory by matter degrees of freedom. The action for these matter degrees of freedom shall  turn out to also be topological-holomorphic in nature. Moreover, we shall explain how the 5d non-commutative Chern-Simons-matter theory describing the conifold is equivalent to 5d non-commutative Chern-Simons theory for a supergroup. 
We shall also analyze general topological-holomorphic 5d Chern-Simons-matter theories that are not necessarily equivalent to 5d CS for a supergroup.

As part of our analysis of 5d Chern-Simons and Chern-Simons-matter actions, we shall study them on the 5-manifold $\mathbb{R}_+ \times \mathbb{C}^2$, with boundary degrees of freedom coupled in a gauge invariant manner. From the perspective of twisted M-theory, one could understand the study of 5d Chern-Simons theory with a boundary as a study of twisted M-theory with a boundary. It is also tempting to interpret the boundary as an M9-brane, although it has been argued that the latter is incompatible with the twist of M-theory \cite{Oh:2021bwi}. It is however, possible that the boundary degrees of freedom describe a domain wall in twisted M-theory. We note that certain domain walls in twisted M-theory have been studied before by Oh and Zhou \cite{Oh:2021bwi}, but are described by a theory of a different nature from the one that we shall encounter here.

In the process of studying 5d Chern-Simons-matter theory with a boundary, we shall generalize the usual 3d-2d correspondence between a 3d Chern-Simons theory on a manifold with a chiral 2d WZW model  \cite{Witten:1988hf} on the boundary of that manifold to a non-commutative 5d-4d correspondence, involving 5d Chern-Simons theory and a novel ``chiral" 4d WZW model. To be precise, the requirement of gauge invariance shall lead us to a gauged, non-commutative ``chiral" 4d WZW model. We emphasize that this is a novel 4d WZW model with a different action from other 4d WZW models previously studied in the literature \cite{Inami:1996zq,Losev:1995cr,Witten:1983tw}. It is, however, related via dimensional reduction to a 3d analogue of the ``chiral" WZW model studied by one of the authors in \cite{Ashwinkumar:2020gxt}.

We note that the relation between 3d Chern-Simons theory and boundary 2d WZW theory has been generalized to the non-commutative case before \cite{Chu:2000bz, Gorsky:2001iq}.  It is thus  interesting to analyze if this correspondence between 3d and 2d systems can be generalized to a 5d-4d correspondence, and we shall show that this is indeed the case.

We shall also study the current algebra associated with the 4d chiral WZW model. We shall compute it first at the classical level, and find that it takes the form of a toroidal current algebra. Moreover, we shall explain how this current algebra arises from radial quantization along one of the complex planes.

We shall also study a 5d-4d correspondence between 5d non-commutative Chern-Simons theory and the 4d chiral WZW model at the quantum level. 
5d non-commutative Chern-Simons theory is in fact power counting non-renormalizable, and to show that it is indeed well-defined as a quantum field theory, at least in perturbation theory, one approach is to use the techniques from BV quantization \cite{Costello:2016nkh}. The extension of the BV formalism to the case with boundaries is known as the BV-BFV formalism, first introduced by Cattaneo, Mnev, and Reshetikhin in \cite{ Cattaneo:2012qu,Cattaneo:2015vsa}.

We shall first employ this formalism to show that a gauged non-commutative 4d chiral WZW model is a holographic dual of 5d non-commutative Chern-Simons theory on $I \times \mathbb{C}^2$  at the quantum level, both for ordinary Lie groups as well as supergroups. The 4d effective action plays the role of a generating functional for the aforementioned toroidal current algebra. Finally, we generalize this duality to 5d non-commutative Chern-Simons-matter theory for the case of adjoint matter, obtaining a 4d chiral WZW model coupled to matter fields in the adjoint representation. This provides us with a computation of the partition function of  5d Chern-Simons-matter theory for adjoint matter, in terms of an effective action consisting of the 4d chiral WZW model coupled to matter fields.

\section{M-theory on a Conifold and 5d Chern-Simons-Matter Theories }

In this section, we shall generalize Costello's description of twisted M-theory in a Taub-NUT background, by considering twisted M-theory on a general conifold. As we shall show, the effective dynamics can be described a 5d topological-holomorphic Chern-Simons-matter theory.

Before proceeding, let us recall the explicit form of 5d non-commutative Chern-Simons theory that describes twisted M-theory in a Taub-NUT background with nontrivial $C$-field. This theory has the action \cite{Costello:2016nkh}  
\begin{equation}\label{act1}
\frac{1}{ \hbar } \int_{\mathbb{R}\times \mathbb{C}^2}  d z \wedge d w \wedge \operatorname{Tr} A \wedge _* d A+\frac{1}{\hbar}\frac{2}{3  } \int_{\mathbb{R}\times \mathbb{C}^2}  d z \wedge d w \wedge\operatorname{Tr}  A \wedge_* A \wedge_* A, 
\end{equation}
where $C\wedge_* D = dx^i\wedge dx^j C_i * D_j$ for one-forms $C$ and $D$,
and where $*$ is the Moyal product that depends on a parameter $\epsilon$, which measures the non-commutativity between the holomorphic coordinates $z$ and $w$ parametrizing the two complex planes $\mathbb{C}_z$ and $\mathbb{C}_w$, that is, we have 
\begin{equation}
f * g=f g+\epsilon \frac{1}{2} \varepsilon_{i j} \frac{\partial}{\partial z_i} f \frac{\partial}{\partial z_j} g+\epsilon^2 \frac{1}{2^2 \cdot 2!} \varepsilon_{i_1 j_1} \varepsilon_{i_2 j_2}\left(\frac{\partial}{\partial z_{i_1}} \frac{\partial}{\partial z_{i_2}} f\right)\left(\frac{\partial}{\partial z_{j_1}} \frac{\partial}{\partial z_{j_2}} g\right)+\ldots
\end{equation}
for elements $f,g$ of the ring of holomorphic functions on $\mathbb{C}_w \times \mathbb{C}_z$, where $z_1=z$ and $z_2=w$, where $\varepsilon_{ij}$ is the alternating symbol and the summation convention is enforced.

Note that due to the presence of the two-form $dz \wedge dw$ wedged with the Chern-Simons three-form in the action, the gauge field here is effectively
a \textit{partial} connection of the form 
\begin{equation}
A=A_tdt   +A_{\bar{z}} d\bar{z} + A_{\bar{w}} d\bar{w} .
\end{equation}
The 5d Chern-Simons theory action is moreover invariant under gauge transformations of the form
\ie 
A_i \rightarrow A_i - \partial_i \alpha - A_i* \alpha + \alpha * A_i,
\fe
where $\alpha$ is a gauge transformation parameter valued in the Lie algebra of the gauge group of the theory.

Let us now proceed to understand how 5d Chern-Simons theory with the action \eqref{act1} generalizes when we replace the Taub-NUT background of twisted M-theory by a general conifold.
\subsection{Twisted M-theory on a Conifold}

 Recall that for $\mathbb{C}^4$ parameterized by complex coordinates $x$, $y$, $u$ and $v$, a general conifold is defined by 
\ie 
x y=v^K u^N.
\fe 
This conifold can be viewed as a $\mathbb{C}^*$ fibration over $\mathbb{C}^2$ parameterized by $v$ and $u$, orbifolded by $\mathbb{Z}_K\times \mathbb{Z}_N$. 

Let us investigate twisted M-theory in this conifold background, which can be understood as two different Taub-NUT manifolds that share two common dimensions, which includes the circle fiber for both Taub-NUTs. As explained in \cite{demello}, shrinking the circle fiber gives us a stack of $N$ D6-branes and a stack of $K$ D6-branes, that are each orthogonal to each other apart from having five common dimensions.

Shrinking the circle fiber that forms the 11th dimension of M-theory leads us to a twist of type IIA string theory.
The twist of type IIA supergravity and string theory was originally studied by Costello and Li \cite{Costello:2016mgj}.
When one has a supergravity background with a Killing spinor that squares to zero, a twist of supergravity can be defined by setting the bosonic ghost to be equal to the square-zero Killing spinor \cite{Costello:2016mgj}.
Moreover, to define supergravity in a twisted $\Omega$-background, one employs a generalized Killing spinor that squares to a rotation, with the twist arising from the identification of this Killing spinor with the ghost field. 

Let us specify the directions transverse to the conifold on which  M-theory is defined. We shall study M-theory on a manifold with $G_2$ holonomy,\footnote{Twisted M-theory on a specific class of manifolds with $G_2$ holonomy was previously studied in \cite{DelZotto:2021ydd,Oh:2022unv}, but this class does not include the $G_2$ manifolds we consider in this paper.} which is the direct product of a general conifold and the real line $\mathbb{R}$. The remaining directions consist of the complex surface $\mathbb{C}^2$. As in \cite{Costello:2016nkh}, we also turn on the 3-form $C$-field background
\ie 
-\epsilon_2 V^{\flat} \wedge d \bar{z} \wedge d\bar{w}.
\fe
where $V^{\flat}$ is the 1-form dual to $V$, the vector field generating the $S^1$ action of the circle fiber of the general conifold. 
Denoting $x^i$ for $i=1,\ldots,11$ as local real coordinates on the 11-dimensional space that M-theory is defined, the following diagram elucidates the M-theory background of interest:
\ie
\overbrace{\hspace{2.1cm}}^{\mathbb{C}^2}
\overbrace{\makebox[0pt]{}}^{\mathbb{R}}
\overbrace{\hspace{3.6cm}}^{\text{General Conifold}}\\
\label{}
\begin{array}{c|c|c|c|c|c| c| c|c|c | c |cc} 
\textrm{M-theory} & x^1 & x^2 & x^3 & x^4 & \textrm{ }x^5\textrm{ }& x^6 & x^7 & x^8 & x^9 & x^{10}  & x^{11}& 
\end{array}
\fe 
Here, the $x^{10}-x^{11}$ directions form the $\mathbb{C}^*$ fiber over the $\mathbb{C}^2$ base of the conifold, with the latter being parametrized by the complex coordinates $v$ and $u$, or equivalently, by the real coordinates $x^6$, $x^7$, $x^8$ and $x^9$. In particular $x^{11}$ is the local coordinate of a circle fiber.

Shrinking the radius of the 11-th dimension parametrized by $x^{11}$, we arrive at the following configuration of $N$ D6-branes and $K$ D6'-branes:
\ie
\overbrace{\hspace{2.1cm}}^{\mathbb{C}^2}
\overbrace{\makebox[0pt]{}}^{\mathbb{R}}
\overbrace{\hspace{1 cm}}^{\mathbb{R}_{\epsilon_1}^2}
\overbrace{\hspace{1 cm}}^{\mathbb{R}_{\epsilon_3}^2}
\overbrace{\hspace{0.3cm}}^{\mathbb{R}}\\
\label{t4}
\begin{array}{c|c|c|c|c|c|c|c|c|c|c}
& x^1 & x^2 & x^3 & x^4 & \textrm{ }x^5\textrm{ } & x^6 & x^7 & x^8 & x^9 & x^{10} \\
\hline 
D6 & \times & \times & \times & \times & \times & \times & \times &  &  &  \\
\hline  
D6' & \times & \times & \times & \times & \times &  &  & \times & \times &  \\
\end{array}
\fe 
where we have considered a twist with two $\Omega$-deformation parameters remaining after the circle reduction, denoted $\epsilon_1$ and $\epsilon_3$.\footnote{The Calabi-Yau condition which is satisfied by the general conifold will impose a relation on the $\Omega$-deformation parameters $\epsilon_1$ and $\epsilon_3$, which is $\epsilon_1+\epsilon_3 =0 $. 
} 
In general, the worldvolume theories of the D6- and D6'-branes will localize to two 5d Chern-Simons theories with gauge groups $U(N)$ and $U(K)$, respectively. Both 5d theories are coupled to a 5d matter action that we will derive in the next subsection. 

However, in what follows, for simplicity we shall focus only on the 5d Chern-Simons theory arising from the D6-branes, and the matter fields that couple to it. In other words, we shall freeze the gauge field degrees of freedom arising from the D6'-branes.
In this case, the positive integer, $K$, if greater than 1, gives rise to a flavour symmetry for the matter action that we will derive below.
Due to the symmetry between two stacks of D6-branes, or equivalently between the two Chern-Simons theories, exactly the same results will hold for the $U(K)$ theory.

\subsection{Matter from Topological Strings}

Let us now derive the matter action by studying the 10d topological string theory that has arose from shrinking the radius of the 11th dimension of twisted M-theory. 
As explained in \cite{Costello:2016nkh}, the twist of type IIA string theory at hand is an A-model
along the  $\mathbb{R}^6$ parametrized by $x^5$ to $x^{10}$ and a B-model along $\mathbb{C}^2$  parametrized by $x^1$ to $x^{4}$,\footnote{Strictly speaking, for $\epsilon_2 \neq 0$, we have a non-commutative B-model along $\mathbb{C}^2$, which is equivalent to an A-model with a non-trivial B-field.} as depicted in the following table:
\ie
\overbrace{\hspace{2.1cm}}^{\text{B-twist}}
\overbrace{\hspace{3.3cm}}^{\text{A-twist}}\\
\label{tmain}
\begin{array}{c|c|c|c|c|c|c|c|c|c|c}
& x^1 & x^2 & x^3 & x^4 & x^5 & x^6 & x^7 & x^8 & x^9 & x^{10} \\
\hline 
D6 & \times & \times & \times & \times & \times & \times & \times &  &  &  \\
\hline  
D6' & \times & \times & \times & \times & \times &  &  & \times & \times &  \\
\end{array}
\fe 
The open strings stretched between the D6-branes and D6'-branes give rise to a matter action, coupled to the gauge fields on each stack of branes. To describe these matter fields, we simply need to identify the space  of supersymmetric states of the aforementioned open strings. 

Instead of doing this directly, we shall use T-duality to take us to a simpler twisted theory, where it is easier to compute the space of supersymmetric states, and then reverse the T-dualities.
T-dualizing along $x^2$ and $x^{4}$, we obtain an A-model on $\mathbb{R}^{10}$ with the brane configuration
\ie
\overbrace{\hspace{5.4cm}}^{\text{A-twist}}\\
\label{t4}
\begin{array}{c|c|c|c|c|c|c|c|c|c|c}
& x^1 & x^2 & x^3 & x^4 & x^5 & x^6 & x^7 & x^8 & x^9 & x^{10} \\
\hline 
D4 & \times &  & \times &  & \times & \times & \times &  &  &  \\
\hline  
D4' & \times & & \times &  & \times &  &  & \times & \times &  \\
\end{array}
\fe  
The crucial point is that the D4- and D4'-branes here are the simplest type of A-branes, i.e., middle-dimensional Lagrangian submanifolds, and the strings stretched between them are described by Lagrangian intersection Floer cohomology.\footnote{With $\Omega$-deformation included along the $x^6-x^7$ and $x^8-x^9$ planes, the $D4$- and $D4'$-brane worldvolume theories each localize to 3d analytically-continued Chern-Simons theory \cite{Luo:2014sva}. }

Let us first focus on the case where non-commutativity is turned off. Following 
Witten's algorithm in \cite{witcsg} we can deduce the matter action arising from the D4-D4' strings, which can be interpreted as an open string field theory action. Recall that from this perspective, the open string states form a differential graded algebra $\mathcal{A}$, where the differential is the BRST operator of the twisted theory. Moreover, in the BV-BRST formalism, the physical fields of the theory will correspond to $\mathcal{A}[0]$, i.e., they have ghost number 0, while ghost fields will have ghost number 1. The supersymmetric ground states of the D4-D4' and D4'-D4 strings ought to be valued in the Lagrangian intersection Floer cohomology for the two branes, which in the present case of D4- and D4'-branes intersecting along $\mathbb{R}^3$ reduces to de Rham cohomology on $\mathbb{R}^3$. The physical states arising from the D4-D4' and D4'-D4 strings, in the case of $N$ D4-branes and $K$ D4'-branes, can be thus modelled, respectively, by the field content 
\ie 
\boldsymbol{\eta} & \in \Omega^*(\mathbb{R}^3)\otimes \textrm{Hom}(\mathbb{C}^N,\mathbb{C}^K)[-1]\\
\boldsymbol{\phi} & \in \Omega^*(\mathbb{R}^3)\otimes \textrm{Hom}(\mathbb{C}^K,\mathbb{C}^N)[-1],
\fe
where the ghost number 0 components of $\boldsymbol{\eta} $ and $\boldsymbol{\phi} $ shall be denoted $\eta$ and $\phi$.

The gauge invariant action governing the dynamics of the ghost number 0 fields takes the form
\ie
\int_{\mathbb{R}^3} \eta \wedge d_A \phi,
\fe
where $\eta$ and $\phi$ are  full one-forms on the three-manifold $\mathbb{R}^3$ parametrized by $x^1$, $x^3$ and $x^5$,
and $d_A$ is the gauge covariant extension of the exterior derivative. The matter fields are in the fundamental representation of the gauge group $U(N)$ and its conjugate, and the fields likewise transform  in the fundamental representation and conjugate fundamental representation with respect to the additional $U(K)$ flavor symmetry which remains after we freeze the gauge field on the $D4'$ worldvolume.

Now, to return to the configuration in \eqref{tmain}, we T-dualize along $x^2$ and $x^4$ again. 
These T-dualities lead us to the 5d topological-holomorphic matter action 
\ie \label{matactt}
\int_{\mathbb{R} \times \mathbb{C}^2} dz\wedge dw \wedge \eta \wedge d_A\phi,
\fe
since T-dualization amounts to making the replacements \cite{yamazaki}
\ie 
D_{x^1} \rightarrow D_{\bar{z}}\\
D_{x^3} \rightarrow D_{\bar{w}},
\fe
where $z$ and $\bar{z}$ parametrize the $x^1-x^2$ plane and  $w$ and $\bar{w}$ parametrize the $x^3-x^4$ plane, and $D_i$ denotes components of the covariant derivative. We emphasize that we have only employed T-duality to a full A-model on 10d space to simplify and elucidate our derivation of \eqref{matactt}. We could have directly computed the space of open string states of topological strings stretching between the D6 and D6'-branes in \eqref{tmain}, which would be 
\ie H F^*\left(\mathbb{R}_{t}, \mathbb{R}_{t}\right) \otimes \operatorname{Ext}^*\left(\mathbb{C}^2, \mathbb{C}^2\right)
\fe 
a direct product of Lagrangian intersection Floer cohomology along the $\mathbb{R}$ parametrized by $x^5$, which we have relabelled as $t$, and the Ext-group for sheaves on $\mathbb{C}^2$, and arrived at the matter action \eqref{matactt}.

Let us consider the case where $K=1$, a restriction we shall make in subsequent sections as well. The explicit form of the 5d topological-holomorphic matter action is 
\ie 
&\int_{\mathbb{R} \times \mathbb{C}^2} dz\wedge dw \wedge \eta \wedge d_A\phi\\=& \int_{\mathbb{R} \times \mathbb{C}^2} dz\wedge dw \wedge dt \wedge  d\bar{z} \wedge d\bar{w} \textrm{ } \varepsilon^{ijk }\eta_{iI} D_j \phi^I_k
\fe
where the antisymmetric tensor $\varepsilon^{ijk}$ is defined such that $\varepsilon^{t\bar{z}\bar{w}}=1$, where
\ie 
\phi = \phi_t dt + \phi_{\bar{z}}d\bar{z} + \phi_{\bar{w}}d\bar{w}\fe
and 
\ie 
\eta &= \eta_t dt + \eta_{\bar{z}} d\bar{z} + \eta_{\bar{w}} d\bar{w}
,
\fe
which transform in a representation $\rho$ of the gauge group and its conjugate $\rho^c$, respectively, and where 
\ie \label{covder}
D_j\phi^I_k &= \partial_j \phi^I_k + A_j^a(T^{\rho}_a)_{\textrm{ }J}^{I}\phi_k^J,
\fe
with $I$ and $J$ are indices for the representations $\rho$ and $\rho^c$. More generally, we can derive gauge invariance for matter with non-commutativity turned on, by generalizing the the covariant derivative \eqref{covder}
to
\ie 
D^*_j\phi^I_k &= \partial_j \phi^I_k + A_j^a(T^{\rho}_a)_{\textrm{ }J}^{I} *\phi_k^J.
\fe
The gauge invariance of this non-commutative matter action is shown in Appendix \ref{matgaug}.
We emphasize that the matter action is gauge invariant by itself, and does not produce any boundary terms when undergoing a gauge transformation.

Thus, starting from twisted M-theory on a conifold, we have obtained an effective description in terms of a 5d Chern-Simons-matter theory of the form 
\ie\label{csm}
&\frac{1}{\hbar } \int_{\mathbb{R}\times \mathbb{C}^2}  d z \wedge d w \wedge \operatorname{Tr} A \wedge_* d A+\frac{1}{\hbar}\frac{2}{3} \int_{\mathbb{R}\times \mathbb{C}^2}  d z \wedge d w \wedge\operatorname{Tr}  A \wedge_* A \wedge_* A  \\&+ \int_{\mathbb{R} \times \mathbb{C}^2} dz\wedge dw \wedge \eta \wedge_* d^*_A\phi
 \\
 = &\frac{1}{\hbar } \int_{\mathbb{R}\times \mathbb{C}^2}  d z \wedge d w \wedge \operatorname{Tr} A \wedge d A+\frac{1}{\hbar}\frac{2}{3} \int_{\mathbb{R}\times \mathbb{C}^2}  d z \wedge d w \wedge\operatorname{Tr}  A \wedge A \wedge_* A  \\&+ \int_{\mathbb{R} \times \mathbb{C}^2} dz\wedge dw \wedge \eta \wedge d^*_A\phi,
\fe
where the equality holds for the integrands up to total derivatives and
\ie 
&\int_{\mathbb{R} \times \mathbb{C}^2} dz\wedge dw \wedge \eta \wedge d^*_A\phi\\=&\int_{\mathbb{R} \times \mathbb{C}^2} dz\wedge dw \wedge dt \wedge  d\bar{z} \wedge d\bar{w} \textrm{ } \varepsilon^{ijk }\eta_{iI} D^*_j\phi^I_k.
\fe
We shall be concerned with this 5d Chern-Simons-matter theory in what follows.

It should be kept in mind that we froze the degrees of freedom arising from the stack of $K$ D6-branes, and set $K=1$ such that there was no remaining flavour symmetry, to arrive at \eqref{csm}. In general, if we do not take these steps, we would have an action involving bifundamental matter fields coupled to dynamical $U(N)$ and $U(K)$ Chern-Simons gauge fields. This action is given explicitly by\footnote{In general there will also be dependence on the Yang-Mills coupling of the D-brane worlvolume theories, which we can remove via a convenient rescaling of the action. }  
\ie \label{fullact}
&\frac{1}{\epsilon_1} \int_{\mathbb{R}\times \mathbb{C}^2}  d z \wedge d w \wedge \operatorname{Tr} A \wedge d A+\frac{1}{\epsilon_1}\frac{2}{3} \int_{\mathbb{R}\times \mathbb{C}^2}  d z \wedge d w \wedge\operatorname{Tr}  A \wedge A \wedge_* A \\
+&\frac{1}{\epsilon_3} \int_{\mathbb{R}\times \mathbb{C}^2}  d z \wedge d w \wedge \operatorname{Tr} A' \wedge d A'+\frac{1}{\epsilon_3}\frac{2}{3} \int_{\mathbb{R}\times \mathbb{C}^2}  d z \wedge d w \wedge\operatorname{Tr}  A' \wedge A' \wedge_* A' 
\\+& \int_{\mathbb{R} \times \mathbb{C}^2} dz\wedge dw \wedge \eta \wedge d^*_{A,A'}\phi,
\fe
where $A'$ denotes the $U(K)$ gauge field, and 
\ie 
&\int_{\mathbb{R} \times \mathbb{C}^2} dz\wedge dw \wedge \eta \wedge d^*_{A,A'}\phi\\=&\int_{\mathbb{R} \times \mathbb{C}^2} dz\wedge dw \wedge dt \wedge  d\bar{z} \wedge d\bar{w} \textrm{ } \varepsilon^{ijk }\eta_{iII'} (\partial_j \phi^{II'}_k + A_j^a(T^{\rho}_a)_{\textrm{ }J}^{I} *\phi_k^{JI'}+ A_j^b(T^{\rho'}_b)_{\textrm{ }J'}^{I'} *\phi_k^{IJ'}),
\fe
where $I'$ and $J'$ are indices associated with a representation of $U(K)$, denoted $\rho'$, under which $\phi$ transforms.

The action \eqref{fullact} can in fact be rewritten as a single 5d Chern-Simons action, but where the gauge group is the  supergroup $U(N|K)$, as in the work of Mikhaylov and Witten \cite{Mikhaylov:2014aoa}, after an appropriate rescaling of the matter fields.
To his end, we recall that the Calabi-Yau condition is $\epsilon_1+\epsilon_3=0$, and set $\epsilon_1=\hbar$ as before.
The action \eqref{fullact} can then be written as a 5d non-commutative Chern-Simons theory for the supergroup $U(N|K)$, with the action
\ie\label{csm}
&\frac{1}{\hbar } \int_{\mathbb{R}\times \mathbb{C}^2}  d z \wedge d w \wedge \operatorname{STr} A \wedge d A+\frac{1}{\hbar}\frac{2}{3} \int_{\mathbb{R}\times \mathbb{C}^2}  d z \wedge d w \wedge\operatorname{STr}  A \wedge A \wedge_* A,  
 \fe
where the gauge field is 
\ie {A}=\left(\begin{array}{cc}{A}_{U(N)} & \eta  \\ \phi & {A}_{U(K)}\end{array}\right).
\fe
 Here, $\textrm{Str}$ is the supertrace on the Lie algebra of $U(N|K)$, that is equal to the ordinary trace when restricted to the Lie algebra of $U(N)$ and the negative of the trace when restricted to the Lie algebra of $U(K)$.
We note that, analogous to the work of Mikhaylov and Witten \cite{Mikhaylov:2014aoa}, the matter fields can be understood to arise from a twisted 5d $\mathcal{N}=1$ hypermultiplet in the bifundamental representation of $U(N) \times U(K)$. In addition, we note that 4d Chern-Simons theory for a supergroup was derived from a closely related system of intersecting D5-branes
in \cite{Ishtiaque:2021jan}.\footnote{In this work, a relation via T-duality is claimed between a system of intersecting D5-branes subject to the twist of \cite{Costello:2018txb} and a D4-NS5 system T-dual to the configuration of Mikhaylov and Witten and subject to the twist of \cite{Ashwinkumar:2019mtj,Ashwinkumar:2018tmm}.}

The relation of the general conifold to supergroup 5d Chern-Simons theory can also be understood as follows. M-theory on a toric Calabi-Yau is dual to type IIB string theory with a $(p,q)$ brane web configuration \cite{Aharony:1997bh,Leung:1997tw}. For the general conifold, this brane configuration would be a set of $N$ D5-branes and another set of $K$ D5-branes ending on a single NS5-brane, in the configuration
\ie
\label{t4}
\overbrace{\hspace{2.1cm}}^{\text{B-twist}}
\overbrace{\hspace{3.3cm}}^{\text{A-twist}}\\
\begin{array}{c|c|c|c|c|c|c|c|c|c|c}
& x^1 & x^2 & x^3 & x^4 & x^5 & x^6 & x^7 & x^8 & x^9 & x^{10} \\
\hline 
D5 & \times & \times & \times & \times & \times & \times^{\uparrow} &  &  &  &  \\
\hline  
D5' & \times & \times & \times & \times & \times & \times^{\downarrow} &  &  &  &  \\
\hline  
NS5 & \times & \times & \times & \times & \times &  &  &  &  &  \times\\
\end{array}
\fe  
where $\uparrow$ and $\downarrow$ indicate the two different half-lines that make up the direction parametrized by $x^6$.
This D5-NS5-D5' configuration is related via T-duality to a D4-NS5-D4' sytem with  topological-holomorphic twist (generalizing the topological-holomorphic  twist of the D4-NS5 system studied in  \cite{Ashwinkumar:2019mtj}) and via a further T-duality to the D3-NS5-D3' system of Mikhaylov-Witten \cite{Mikhaylov:2014aoa}. The D5-NS5-D5' system thus localizes to 5d Chern-Simons theory for the supergroup. 
The relationship between the conifold and the affine Yangian for supergroups has been predicted before, see, for example, \cite{Li:2020rij} and \cite{Butson:2023eid}.
Notably, this suggests that the Gromov-Witten invariants of the conifold are associated in some manner with the affine Yangian for supergroups.

It shall be useful to compare the 5d Chern-Simons-matter action derived here to the
 3d Chern-Simons-matter theory defined on a three-manifold with transverse holomorphic foliation, that  was first studied by Aganagic, Costello, McNamara and Vafa \cite{Aganagic}. Notably, this theory can be interpreted as the theory on a   
Lagrangian 3-brane in an A-type topological string theory on a Calabi-Yau 3-fold, in the presence of a coisotropic 5-brane. The action of the model has the form
\begin{equation}\label{acmv}
S=\frac{i k}{4 \pi} \int_M C S(A)+\int_M \tilde{\eta} \wedge d_A \tilde{\phi},
\end{equation}
where $M$ is a 3-manifold endowed with a transverse holomorphic foliation, where $\tilde{\phi} \in \Omega^{1,0} \otimes R $  and $\tilde{\eta} \in \Omega / \Omega^{1,0} \otimes R^c$ are the matter fields (where ``1,0" refers to the holomorphic direction of $M$), where $R$ is the representation of the gauge group, $G$, and $R^c$ is the conjugate representation.  
The matter term in this action can be written explicitly as 
\begin{equation}\label{holmat}
\int_M\left(\tilde{\eta}_{\bar{u}} D_t \tilde{\phi}_u-\tilde{\eta}_t D_{\bar{u}} \tilde{\phi}_u\right) d t \wedge d u \wedge d \bar{u},
\end{equation}
where $u$ and $\bar{u}$ are complex coordinates in the holomorphic direction of $M$, and $t$ is the coordinate parametrizing the remaining direction, and where the following covariant derivatives are employed: 
\begin{equation}
\begin{aligned}
    D_t\tilde{\phi}^I_u &= \partial_t \tilde{\phi}^I_u + A_t^a(T^R_a)_{\textrm{ }J}^{I}\tilde{\phi}_u^J,\\
        D_{\bar{u}}\tilde{\phi}^I_u &= \partial_{\bar{u}} \tilde{\phi}^I_u + A_{\bar{u}}^a(T^R_a)_{\textrm{ }J}^{I}\tilde{\phi}_u^J,
    \end{aligned}
\end{equation}
where $I$ and $J$ are indices for the representations $R$ and $R^c$. The main difference between this 3d Chern-Simons-matter theory and the 5d Chern-Simons-matter theory is that 3d Chern-Simons theory is a topological theory, and the introduction of the matter described above breaks this symmetry. On the other hand, 5d Chern-Simons theory is inherently topological-holomorphic. Indeed, as we have seen, the dimensional reduction of the latter to 3d produces a topological 3d matter action, that differs from the 3d matter action in \eqref{holmat}.

\section{Boundaries in M-theory and the Gauged Chiral 4d WZW Model }

Having described  twisted M-theory on a conifold in terms of a general conifold background in terms of a Chern-Simons-matter theory, we would like to understand how this system can be generalized to include boundaries in 11-dimensional space, in particular, along the $x^5$ direction, which we shall parametrize by the coordinate $t$. 

Recall that the relation between 3d topological field theories defined on a three-manifold and 2d conformal field theories on the boundary, especially the 3d Chern-Simons/2d Wess-Zumino-Witten correspondence, is well known \cite{Witten:1988hf}. For Chern-Simons theory on a manifold with a boundary, since the gauge field appears linearly in action, a component thereof subject to a Dirichlet boundary condition can be shown to be auxiliary, and thus integrated out, and this in turn imposes a constraint that a component of the field strength vanishes. Solving this constraint results in a 2d WZW model \cite{Moore:1989yh,Elitzur:1989nr}. The Chern-Simons degrees of freedom manifest as edge modes described by the boundary WZW model in this correspondence.

Alternatively, instead of imposing boundary conditions on the Chern-Simons gauge field, it is possible to introduce new boundary degrees of freedom from the start to preserve (all or part of) the bulk gauge symmetry, resulting in a coupled 2d boundary action, such that the coupled 3d-2d system is gauge invariant \cite{Chu:2009ms, Berman:2009xd}. This is particularly important when the Chern-Simons theory is coupled to bulk matter fields where boundary conditions cannot be used to explicitly obtain a 3d-2d correspondence. This formalism has been used to study M2-branes (described by the ABJM model \cite{Aharony:2008ug}) ending on M5-branes \cite{Chu:2009ms, Berman:2009xd, Faizal:2016skd, Faizal:2011cd}.

In order to investigate the 5d Chern-Simons-matter action in the presence of a boundary, it is likewise not natural to employ boundary conditions to obtain a boundary dual, due to the presence of the interaction term coupling the matter fields to the gauge fields.
Our goal in this section is thus to couple a 4d boundary theory to the 5d Chern-Simons matter action in a gauge invariant manner.

To this end, we shall first introduce a novel gauged 4d WZW model. As we shall see, the terms arising from the gauge transformation of the gauged 4d WZW action cancel the extra terms that appear in the gauge-transformed 5d CS action. 
This 4d theory is an analogue of the 2d chiral WZW model, and we shall thus refer to it as the 4d ``chiral" WZW model.
The gauged ``chiral" 4d WZW action has the form
\begin{eqnarray} \label{action00}
  S^{4d,gauged}_{WZW}[A, G]& =&S^{4d}_{WZW}[G] - 2\int_{\mathbb{C}^2 } dz\wedge dw \wedge d \bar{z} \wedge d\bar{w} \textrm{Tr} (A_{\bar{w}} \partial_{\bar{z}} G G^{-1}) \nonumber\\
  &-&\int_{\mathbb{C}^2 }  dz\wedge dw \wedge d \bar{z} \wedge d\bar{w} \textrm{Tr} (A_{\bar{z}}A_{\bar{w}})
\end{eqnarray} 
where 
\begin{equation} 
\begin{aligned}
 S^{4d}_{WZW}[G] =&- \int_{\mathbb{C}^2 }  dz\wedge dw \wedge d \bar{z} \wedge d\bar{w} \textrm{Tr}\left( G^{-1} \partial_{\bar{z}}G G^{-1} \partial_{\bar{w}}G\right) \\&-\frac{1}{3}\int_{\mathbb{R}_+\times \mathbb{C}^2 } dz \wedge dw \wedge \textrm{Tr}(G^{-1}dG\wedge G^{-1}dG\wedge G^{-1}dG ).
\end{aligned}
\end{equation}

In what follows, we shall analyze the gauge transformation of this action in detail.

\subsection{Gauge Transformations  of the 4d Chiral WZW Model }
\label{finitegauge}

The gauge transformation of the gauge field is defined as 
\begin{equation}
A = -dgg^{-1}+g \tilde{A} g^{-1}
\end{equation}
 while the gauge transformation of $G$ is defined as $ 
G = g \tilde{G}
$.
Under this gauge transformation, the action \eqref{action00} transforms to
\begin{eqnarray} \label{trwz}
 S^{4d, gauged}_{WZW}[A, G]= &&S^{4d}_{WZW}[g\tilde{G}] - 2\int_{\mathbb{C}^2 } dz\wedge dw \wedge d \bar{z} \wedge d\bar{w} \textrm{Tr} (g\tilde{A}_{\bar{w}}g^{-1}-\partial_{\bar{w}}gg^{-1}) \partial_{\bar{z}} (g\tilde{G}) (g\tilde{G})^{-1}) \nonumber\\& -&\int_{\mathbb{C}^2 }  dz\wedge dw \wedge d \bar{z} \wedge d\bar{w} \textrm{Tr} ((g\tilde{A}_{\bar{z}}g^{-1}-\partial_{\bar{z}}gg^{-1})(g\tilde{A}_{\bar{w}}g^{-1}-\partial_{\bar{w}}gg^{-1})).
\end{eqnarray} 

Using the Polyakov-Wiegmann identity,
\begin{eqnarray} 
S_{WZW}^{4d}[g\tilde{G}] = S_{WZW}^{4d}[g] +S_{WZW}^{4d}[\tilde{G}] - 2 \int_{\mathbb{C}^2} d^4x \textrm{Tr} (g^{-1}\partial_{\bar{w}}g\partial_{\bar{z}}\tilde{G}\tilde{G}^{-1}),
\end{eqnarray}
where we have defined $d^4x= dz\wedge  dw \wedge d\bar{z}\wedge  d\bar{w} $,
the expression \eqref{trwz} can be rewritten as 
\begin{equation} 
\begin{aligned}
 S^{4d, gauged}_{WZW}[A, G]\\=  S_{WZW}^{4d}[g] +S_{WZW}^{4d}[\tilde{G}]\nonumber \\- 2 \int_{\mathbb{C}^2} d^4x \textrm{Tr} \Big(&g^{-1}\partial_{\bar{w}}g\partial_{\bar{z}}\tilde{G}\tilde{G}^{-1} + g \tilde{A}_{\bar{w}}g^{-1}\partial_{\bar{z}}gg^{-1}+ g \tilde{A}_{\bar{w}}g^{-1}g\partial_{\bar{z}}\tilde{G}\tilde{G}^{-1}g^{-1}\nonumber \\&-\partial_{\bar{w}}gg^{-1}\partial_{\bar{z}}gg^{-1}-\partial_{\bar{w}}gg^{-1}g\partial_{\bar{z}}\tilde{G}\tilde{G}^{-1}g^{-1} + \frac{1}{2} \tilde{A}_{\bar{w}}\tilde{A}_{\bar{z}}- \frac{1}{2} g\tilde{A}_{\bar{z}}g^{-1}\partial_{\bar{w}}gg^{-1}\nonumber\\ &- \frac{1}{2} \partial_{\bar{z}}gg^{-1}g\tilde{A}_{\bar{w}}g^{-1}+ \frac{1}{2} \partial_{\bar{z}}gg^{-1}\partial_{\bar{w}}gg^{-1}\Big).
 \end{aligned}
\end{equation}
The first term in the parentheses cancels the fifth term in the parentheses, while the terms that purely depend on $g$ cancel the kinetic term of $ S_{WZW}^{4d}[g]$.  The remaining terms can be reexpressed as
\ie \label{gaugetx}
 & S^{4d,gauged}_{WZW}[\tilde{A}, \tilde{G}] \\ &+ \int_{\mathbb{R}_+ \times \mathbb{C}^2 } dz \wedge dw \wedge \left(-\textrm{Tr }d (g^{-1} dg \wedge \tilde{A})  - \frac{1}{3}\textrm{Tr} (g^{-1} dg \wedge g^{-1} dg \wedge g^{-1} dg  )\right).
\fe
Now we will show that the finite gauge transformation of this WZW action can cancel that of the bulk 5d CS action if this WZW theory is put on the boundary of the manifold with CS action in the bulk. 

To demonstrate that, we  would like to compute the variation of the 5d CS action 
\begin{eqnarray} 
\int_{\mathbb{R}_+ \times \mathbb{C}^2} dz \wedge dw \wedge  CS(A).
\end{eqnarray} 
We shall consider the gauge variation of the Chern-Simons functional
under finite gauge transformations of the form 
$ 
A =-dgg^{-1}+g \tilde{A} g^{-1}.
$

To compute this, we define
$
\widehat{A}:=-d {g} {g}^{-1}, \quad A^{\prime}:={g} \tilde{A} {g}^{-1}.
$
For a sum of two connections, the Chern-Simons functional can be written as 
\begin{eqnarray}
\begin{aligned}
C S(A) & =\int \textrm{Tr} \bigg((\widehat{A}+A')\wedge d(\widehat{A}+A')+\frac{2}{3}(\widehat{A}+A')\wedge (\widehat{A}+A')\wedge(\widehat{A}+A') \bigg)\\
&={CS}(\widehat{A})+2 \operatorname{Tr}\left(F(\widehat{A}) A^{\prime}\right)-\mathrm{d} \operatorname{Tr}\left(\widehat{A} A^{\prime}\right)+2 \operatorname{Tr}\left(\widehat{A} A^{\prime} A^{\prime}\right)+{CS}\left(A^{\prime}\right),
\end{aligned} \label{csga}
\end{eqnarray}
where we have integrated by parts in the second term of the second line. 
Since $\widehat{A}$ is pure gauge, it is a flat connection and $F(\widehat{A})=0$, which implies that the second term of the final expression in \eqref{csga} is zero. 
Therefore the gauge variation of the 5d CS action is 
\begin{equation}
    \begin{aligned}
    \label{gaugeg2}
&\int_{\mathbb{R}_+ \times \mathbb{C}^2 } dz \wedge dw \wedge CS(A) \\&=\int_{\mathbb{R}_+ \times \mathbb{C}^2 } dz \wedge dw \wedge \left( CS(\tilde{A})+ \textrm{Tr }d (g^{-1} dg \wedge \tilde{A}) + \frac{1}{3}\textrm{Tr} (g^{-1} dg \wedge g^{-1} dg \wedge g^{-1} dg  )\right).
\end{aligned}
\end{equation}

Thus, since the $g$-dependent terms in \eqref{gaugetx} and \eqref{gaugeg2} cancel, we find that this gauge transformation can be cancelled by including a boundary gauged WZW action that transforms nontrivially under gauge transformations.
\subsection{M2-branes as Local Operators}
A natural class of operators in 5d Chern-Simons theory are Wilson lines along the topological direction. These are identifed with M2-branes from the perspective of twisted M-theory. In the present setup with a boundary, these Wilson lines will end at a finite point, and one needs to be careful in defining the operator in order to preserve gauge invariance.

This leads us to naturally consider \textit{composite} operators. These are
Wilson lines ending on  local operators, defined by the group-valued field $g$, at the boundary of $\mathbb{R}_+$, such that the nontrivial gauge transformations of the Wilson line and the operator $g$ cancel each other, and the composite operator is gauge invariant.

Recall that a Wilson line along a curve $\mathcal{C}$, starting at $t_i$ and ending at $t_f$, and in a representation $R$, satisfies the gauge transformation property
\begin{equation}
\mathcal{P} e^{\int_{\mathcal{C}} A}=g_R(t_i)\mathcal{P} e^{\int_{\mathcal{C}} A'}g^{-1}_R(t_f),
\end{equation}
where $A=g A'g^{-1}-dgg^{-1}$. 
Let us identify $t_f$ as $t=0$ and $t_i$ as $t=\infty$ in the present context. If we specify boundary conditions at $t=\infty$ where all gauge fields go to zero, this means that gauge transformations also ought to be trivial there. Hence we find that $g_R(t_i)$ is the identity element of the gauge group.

In order to preserve gauge invariance at $t=0$, we need to form a composite operator of the form
\ie \label{comp}
\mathcal{P} e^{\int_{\mathcal{C}} A}\cdot G_R(t=0),
\fe 
where $R$ indicates that the group-valued field $G$ is chosen to be in the representation $R$.
Since $G$ transforms via left multiplication by $g$, we find that the composite operator \eqref{comp} is gauge invariant. 
This further leads us to deduce that M2-branes (Wilson lines in 5d CS) can be identified with local operators in the 4d gauged chiral WZW model. We expect to be able to compute correlation functions of M2-branes via correlation functions of local operators in the 4d boundary theory.
\subsection{Non-commutative Deformation  }

We shall now show that a non-commutative generalization of the 5d-4d bulk-boundary system discussed previously is gauge invariant. 
In this case,  we consider the non-commutative 5d Chern-Simons action on $\mathbb{R}_+\times \mathbb{C}_z\times \mathbb{C}_w$ with action of the form \eqref{act1}, together with a
gauged, non-commutative generalization of the 4d chiral WZW model on the boundary, which has the form : 

\begin{eqnarray} 
  S^{4d,gauged}_{NCWZW}[G] =S^{4d}_{NCWZW}[G] - 2\int_{\mathbb{C}^2 } dz\wedge dw \wedge d \bar{z} \wedge d\bar{w} \textrm{Tr} (A_{\bar{w}} *\partial_{\bar{z}} G * G^{-1}) \nonumber\\ -\int_{\mathbb{C}^2 }  dz\wedge dw \wedge d \bar{z} \wedge d\bar{w} \textrm{Tr} (A_{\bar{z}} * A_{\bar{w}})
\end{eqnarray} 
where 
\begin{eqnarray} 
 S^{4d}_{NCWZW}[G] &=- \int_{\mathbb{C}^2 }  dz\wedge dw \wedge d \bar{z} \wedge d\bar{w} \textrm{Tr}\left( G^{-1} * \partial_{\bar{z}}G *G^{-1} *\partial_{\bar{w}}G\right)\nonumber\\&-\int_{\mathbb{R}_+\times \mathbb{C}^2} dz \wedge dw  \wedge \textrm{Tr}(G^{-1} * dG\wedge_* G^{-1}*dG\wedge_* G^{-1}* dG ).
\end{eqnarray}
The (finite) gauge transformations are now of the form
$ 
G =g * \tilde{G}
$ 
and  $
A =-dg*g^{-1}+g *\tilde{A} * g^{-1}, $ 
which generalizes the gauge transformations for trivial non-commutativity.

Let us focus on the $U(N)$ case and list some useful properties of the star product. Firstly, it is associative. A cyclic property reminiscent of a trace also holds when integrated
\begin{eqnarray} 
\int d^5x \textrm{Tr} (A * B )= \int d^5x \textrm{Tr} (A  B) =\int d^5x \textrm{Tr} (B A )= \int d^5x \textrm{Tr} (B * A)
\end{eqnarray} 
It is also straightforward to define a group element of $U(N)_*$ via $g=e^{\phi}=1+ \phi + \frac{1}{2} \phi * \phi + \frac{1}{3!} \phi * \phi * \phi + \ldots $, where $\phi \in \mathfrak{u}(N)$. Its inverse defined via $g * g^{-1} =1$ is then 
\begin{eqnarray} 
g^{-1}=e^{-\phi}=1 - \phi + \frac{1}{2} \phi * \phi - \frac{1}{3!} \phi * \phi * \phi + \ldots 
\end{eqnarray} 

We can also define the Maurer-Cartan form in the usual way, i.e., $-dg * g^{-1}$. Since $g * g^{-1}=1$, we can derive that
$
d(g^{-1}) = - g^{-1} * dg * g^{-1}
$
The following analog of the Maurer-Cartan equation then holds : 
$
d(-dg *g^{-1}) + (-dg *g^{-1})\wedge_* (-dg *g^{-1})=0
$

Using these properties, we find that the derivation of the gauge transformation of the bulk 5d CS action is analogous to that in the commutative case shown in \ref{finitegauge}, with the result
\ie
&\int_{\mathbb{R}_+ \times \mathbb{C}^2 } dz \wedge dw \wedge CS(A)\\=&\int_{\mathbb{R}_+ \times \mathbb{C}^2 } dz \wedge dw \wedge (  CS(\tilde{A})+ \textrm{Tr }d (g^{-1} * dg \wedge_* \tilde{A})\\ & \quad \quad \quad \quad \quad \quad \quad \quad \quad \quad + \frac{1}{3}\textrm{Tr} (g^{-1} * dg \wedge_* g^{-1} * dg \wedge_* g^{-1} * dg  )).
\fe
This is canceled by the gauge transformation of the boundary non-commutative 4d chiral WZW model. 

\subsection{Boundary Matter Action}

We have so far discussed a gauge invariant coupling between 5d non-commutative Chern-Simons theory and a boundary action on the boundary. Since the bulk matter action is gauge invariant regardless of the presence of boundaries, we do not need to include a boundary action involving the matter fields to preserve gauge symmetry. 

In general, when we have a coupled bulk-boundary system, the boundary action can include terms depending on the boundary values of bulk fields. In particular, if we have bulk matter fields there can be a boundary matter action or potential. The form of this boundary matter action may be constrained by symmetries. For example, in the case of an ABJM-like bulk theory the boundary matter action would be a quartic potential as constrained by (classical) conformal symmetry \cite{Chu:2009ms} but taking into account supersymmetry would further constrain the form of this potential \cite{Berman:2009kj}.

In the present case, one can introduce a boundary matter action of the form 
\ie 
c\int_{\mathbb{C}^2} dz \wedge dw \wedge \eta {\wedge_*} \phi,
\fe
where $c$ is a normalization constant. This action satisfies the constraints of being holomorphic (in the sense of only depending on the complex structure of the boundary) as well as gauge invariance.

\subsection{Equivalence of Boundary Action and Boundary Conditions}

While we have not imposed boundary conditions explicitly for the gauge fields, choosing instead to include a boundary action to satisfy gauge invariance, one can show as in \cite{Chu:2009ms} the equivalence of this boundary action to certain boundary conditions on the gauge field. This is particularly simple when the matter fields are not present. For example, in the case of 3d Chern-Simons theories, there is a well-known equivalence to 2d WZW models when there is a boundary and this can be derived by imposing boundary conditions on the gauge field \cite{Elitzur:1989nr} or through coupling to boundary degrees of freedom and then imposing the boundary equations of motion \cite{Chu:2009ms}.
Indeed, varying $A_{\bar{w}}$ gives rise to the boundary condition 
\ie 
A_{\bar{z}}= -\partial_{\bar{z}}G * G^{-1}|_{t=0},
\fe 
which is the analogue of pure gauge for the gauge field of a non-commutative gauge theory.

\section{Current Algebra of the 4d 
Chiral WZW Model}
In the analysis that follows where we derive the current algebra of the 4d chiral WZW model, we shall not be concerned with the matter action, but we shall instead focus on identifying the current algebra associated with the 4d chiral WZW model. 

\subsection{Classical Current Algebra}
Let us derive the current algebra of the 4d chiral WZW model defined on a product of two cylinders, with the time direction identified with the radial direction along the $\mathbb{C}^{\times}$ that one obtains from a conformal transformation of the cylinder parametrized by $(w,\bar{w})$.  To this end, we write 
$w = t + i\sigma_w$, 
where $\sigma_w$ is the compact direction of the cylinder, and $t\in \mathbb{R}$ is identified as time.

The relevant sector of the action \eqref{action00} is first order in time, since we know that the partial derivative $\partial_{\bar{w}}$ appears only once, and that it can be written explicitly as 
\ie 
\partial_{\bar{w}} = \frac{1}{2}(\partial_t - i \partial_{\sigma_w}).
\fe
Since the term that depends on $\partial_{\sigma_w}$ has no time derivatives it  can be ignored in computing Poisson brackets. The sector that contains time derivatives is of the general form
\begin{equation} 
I =\int dt \lambda_i(\phi) \frac{d\phi^i}{d t}.
\end{equation}
Varying this action gives rise to
\begin{equation}
\delta I=\int d t \omega_{i j} \delta \phi^i \frac{d \phi^j}{d t}
\end{equation}
where $\omega_{i j}=\frac{\partial}{\partial \phi^i} \lambda_j-\frac{\partial}{\partial \phi^j} \lambda_i$ are the components of the symplectic form on the classical phase space. We can then obtain the Poisson bracket of any two functions $X$ and $Y$ on the phase space via
\begin{equation}
[X, Y]_{P B}=\omega^{i j} \frac{\partial X}{\partial \phi^i} \frac{\partial Y}{\partial \phi^j}
\end{equation}
where $\omega^{j k} \omega_{k l}=\delta_l^j$.
For the 4d chiral WZW model, we find that the phase space symplectic structure is given by $\omega=1_{\mathfrak{g}} \otimes \frac{(-2)}{\hbar } \frac{\partial}{\partial \bar{z}} \otimes 1_{\sigma_{w}}$, where $1_{\mathfrak{g}}$ acts on the Lie algebra index, $\frac{(-2)}{\hbar} \frac{\partial}{\partial \bar{z}}$ acts on the $(z,\bar{z})$ coordinates, and $1_{\sigma_w}$ acts on the $\sigma_w$ coordinate. Therefore, its inverse is
\begin{equation}
\omega^{-1}=1_{\mathfrak{g}} \otimes\left(-\frac{\hbar}{2}\right)\left(\frac{\partial}{\partial \bar{z}}\right)^{-1} \otimes 1_{\sigma_{w}} .
\end{equation}

Let us now compute the Poisson bracket of $X=\operatorname{Tr} A \frac{\partial g}{\partial \bar{z}} g^{-1}$ and $Y=\operatorname{Tr} B \frac{\partial g}{\partial \bar{z}^{\prime}} g^{-1}$ where $A$ and $B$ are arbitrary generators of the group $G$. This can be done by evaluating $\delta X \delta Y=\frac{\partial X}{\partial \phi^i} \frac{\partial Y}{\partial \phi^j} \delta \phi^i \delta \phi^j$, and subsequently replacing $\delta \phi^i \delta \phi^j$ by $\omega^{i j}$. Proceeding in this manner, we find (with $x$ denoting the spatial coordinates of the 4d manifold on which the WZW model is defined)
\begin{equation}
\delta X \delta Y=\operatorname{Tr} g^{-1}(x) A g(x) \frac{\partial}{\partial \bar{z}}\left(g^{-1} \delta g(x)\right) \cdot \operatorname{Tr} g^{-1}\left(x^{\prime}\right) B g\left(x^{\prime}\right) \frac{\partial}{\partial \bar{z}^{\prime}}\left(g^{-1} \delta g\left(x^{\prime}\right)\right) \text {. }
\end{equation}

To obtain the Poisson bracket, we ought to replace $\left(g^{-1} \delta g(x)\right)^a\left(g^{-1} \delta g\left(x^{\prime}\right)\right)^b$ (where $a$ and $b$ are Lie algebra indices) by
\begin{equation}
\delta^{a b}\left(-\frac{\hbar}{2} \right) \frac{-1}{2\pi i} \frac{1}{z-z'}\delta\left(\sigma_w-\sigma_w^{\prime}\right)
\end{equation}
where $ \frac{-1}{2\pi i} \frac{1}{z-z'}$ is the inverse of $\frac{\partial}{\partial \bar{z}}$. Therefore, $\frac{\partial}{\partial \bar{z}}\left(g^{-1} \delta g(x)\right)^a \cdot \frac{\partial}{\partial \bar{z}^{\prime}}\left(g^{-1} \delta g\left(x^{\prime}\right)\right)^b$  ought to be replaced by $\delta^{a b}\left(- \frac{\hbar}{2}\right) \partial_{\bar{z}}\delta^{(2)}\left(z-z^{\prime}\right)\delta\left(\sigma_w-\sigma_w^{\prime}\right)$. Hence, we arrive at the Poisson bracket
\begin{eqnarray} \label{curalgcl}
\begin{aligned}
{[X, Y]_{P B} } & = \frac{\hbar}{2} \partial_{\bar{z}}\delta^{(2)}\left(z-z^{\prime}\right) \delta\left(\sigma_w-\sigma_w^{\prime}\right) \operatorname{Tr} g^{-1}(\phi, z) A g(\phi, z) g^{-1}\left(\phi^{\prime}, z^{\prime}\right) B g\left(\phi^{\prime}, z^{\prime}\right) \\
& = \frac{\hbar}{2} \delta\left(\sigma_w-\sigma_w^{\prime}\right) \delta^{(2)}\left(z-z^{\prime}\right) \operatorname{Tr}\left([A, B] \frac{\partial g}{\partial \bar{z}} g^{-1}\right)+\frac{\hbar}{2} \delta\left(\sigma_w-\sigma_w^{\prime}\right) \partial_{\bar{z}}\delta^{(2)}\left(z-z^{\prime}\right) \operatorname{Tr} A B
\end{aligned}
\end{eqnarray}
Upon rescaling both sides of this equation by $\left(\frac{-2}{ \hbar}\right)^2$, we arrive at a classical current algebra. Note that we have found a central extension term at the classical level. This is analogous to the case of the standard 2d WZW model with kinetic term in ``chiral" form, i.e., using lightcone coordinates, wherein such a central extension term also appears in the classical Poisson bracket involving currents when taking one of the lightcone directions to be time.

\subsection{Quantum Current Algebra }
We would now like to investigate the quantum current algebra of the 4d chiral WZW model. Let us consider the simplest case of the $U(1)$ chiral 4d WZW model, with the action 

\begin{equation}
\frac{1}{\hbar}\int_{\mathbb{C}^2} d^4x \partial_{\bar{z}} \phi \partial_{\bar{w}}\phi.
\end{equation}
Using the complex analysis lemmata 
\ie 
 \frac{1}{2 \pi i} \bar{\partial}\frac{1}{z}dz&=-\delta^{(2)}(z, \bar{z}) dz \wedge d\bar{z}\\  \frac{1}{2 \pi i} \partial \frac{1}{\bar{z}}d\bar{z}&=-\delta^{(2)}(z, \bar{z})dz \wedge d\bar{z},\fe 
the propagator can then be shown to be 
\ie 
\left\langle\phi(x) \phi(y)\right\rangle=\hbar \Delta(x-y)
\fe
where $x,y$ denote all four complex coordinates, and where 
\ie \Delta(x)= \frac{1}{(2 \pi i)^2} \frac{1}{z}\frac{1}{w}.\fe

In what follows, we shall use a UV regularized version of the  propagator \ie \Delta(x-x') = \frac{1}{(2\pi i )^2}\frac{1}{z-z'}\frac{1}{w-w'},\fe  based on the Feynman parametrization
\ie
\frac{1}{z}=\int_{t=0}^{\infty} \bar{z} e^{-\frac{z \bar{z}}{t}} \frac{\mathrm{d} t}{t^2}.
\fe
We shall regularize both potentially divergent factors $\frac{1}{z-z'}$ and $\frac{1}{w-w'}$ in this manner.
The UV regularization follows from evaluating the integral over the 
region $t\geq \epsilon$, i.e.,
we have 
\ie
\int_{t=\epsilon}^{\infty} \bar{z} e^{-\frac{z \bar{z}}{t}} \frac{\mathrm{d} t}{t^2}= \frac{1-e^{-\frac{z \bar{z}}{\epsilon}}}{z}.
\fe
The regularized propagator then takes the form 
\ie \Delta_{\textrm{reg}}(x-x') = \frac{1}{(2\pi i )^2}\frac{(1-e^{-\frac{|z-z'|^2}{\epsilon}})}{z-z'}\frac{(1-e^{-\frac{|w-w'|^2}{\epsilon}})}{w-w'},\fe 
In particular, in the limit where $z$ approaches $z'$ or $w$ approaches $w'$, the regularized propagator is not divergent, unlike the unregularized one.

Defining 
\ie 
J_{\bar{z}}=\partial_{\bar{z}}\phi
\fe
we can then immediately derive the current algebra 
\ie \label{jjcor}
\langle J_{\bar{z}}(z,w)J_{\bar{z}}(z',w')\rangle = -\hbar \frac{1}{2\pi i} \partial_{\bar{z}}\delta^{(2)}(z-z')\frac{1}{w-w'}.
\fe
Straightforwardly generalizing the usual rules of radial quantization on the complex $w$-plane familiar from 2d CFT, one finds that this current algebra agrees with the classical result \eqref{curalgcl}. To see this explicitly, we employ the following contour integral definition of the commutator 
\ie
[A(z'), B(z)]=\oint_0 d w \oint_w d w' a(z',w') b(z,w),
\fe
for 
\ie 
A(z)&=\oint_0 dw \textrm{ }a(z,w) \\
B(z)&=\oint_0 dw \textrm{ }b(z,w), 
\fe
which gives us 
\ie 
[ J^m_{\bar{z}}(z),J^n_{\bar{z}}(z')] = -\hbar \frac{1}{2\pi i }\partial_{\bar{z}}\delta^{(2)}(z-z')\delta^{m+n,0}
\fe
when applied to the correlator \eqref{jjcor}, where $J^m_{\bar{z}}(z)= \frac{1}{2\pi i} \oint_0 J_{\bar{z}}(z,w) w^{m-\frac{1}{2}}dw$. This is the quantization of the algebra \eqref{curalgcl}, for the case of gauge group $U(1)$. This can be checked by writing the currents and delta function in \eqref{curalgcl} as Fourier series in the periodic coordinates $\sigma_w$ and $\sigma'_w$, and rescaling the currents appropriately. 
 
 For non-commutative U(1) 4d WZW, we can write down the interaction term at leading order in noncommutativity parameter, and compute how the propagator is modified by self-interaction terms, and then compute the corrected current algebra. The interaction terms in the Lagrangian take the form 
\ie 
\frac{\epsilon}{2}\partial_{\bar{z}}(\partial_z \phi \partial_w \phi)\partial_{\bar{w}}\phi + \frac{\epsilon}{2}\partial_{\bar{w}}(\partial_z \phi \partial_w \phi)\partial_{\bar{z}}\phi,
\fe
where $\epsilon$ is the non-commutativity parameter. 
From here, we can read off the Feynman rules that enter the higher loop corrections to the propagator, and compute the corrections to the current algebra. However, it can be shown at least at one-loop, that the current algebra is not modified due to non-commutativity.

A symmetry argument for this observation that the two-point correlation functions do not depend on $\epsilon$ at order $\epsilon$, but are expected to depend at order $\epsilon^2$, is as follows. The interaction terms at order $\epsilon$ are cubic in $\phi$ (and its derivatives) so are invariant under simultaneously changing the sign of $\phi$ and $\epsilon$. However, $J_{\bar{z}}$ depends linearly on (a derivative of) $\phi$ so the two point function is invariant under $\phi \to - \phi$. Therefore the two-point function must also be invariant under a change in sign of $\epsilon$, hence expanding in powers of $\epsilon$ we cannot have any odd powers of $\epsilon$.

\section{5d Chern-Simons-4d Chiral WZW Correspondence from BV-BFV Quantization}

 We shall now study the  quantum 5d non-commutative CS theory on $I \times \mathbb{C}^2$, where $I=[0,1]$, using the BV-BFV formalism. This is a natural approach to the quantization of 5d Chern-Simons theory in the presence of a boundary, since quantization without a boundary is well-described by the BV-formalism in \cite{Costello:2016nkh}. 
 Although we shall not explicitly generalize the proof that 5d Chern-Simons theory is well-defined in perturbation theory to the case with the boundary conditions picked in this section, it is expected that the arguments of Costello in Section 9 and Appendix B of \cite{Costello:2016nkh} hold in this case. 

The computation that follows furnishes a 5d-4d correspodence for twisted M-theory with boundaries. In particular, if the gauge group is $GL(N)$, we are studying M-theory on a Taub-NUT manifold, while if the gauge group is the supergroup $GL(N|K)$, we would be studying the general conifold.

In what follows, we shall impose the boundary conditions 
\ie 
A_{\bar{w}}&=\textrm{constant}|_{t=0},\\
A_{\bar{z}}&=\textrm{constant}|_{t=1},
\fe
which are the  analogues of antiholomorphic to holomorphic boundary conditions of 3d Chern-Simons theory used in \cite{Cattaneo:2020lle}, with the further inclusion of the boundary action
\ie 
S_{boundary}=&\int_{\{1\}\times \mathbb{C}^2} dz \wedge dw \wedge d\bar{z} \wedge d\bar{w}\textrm{Tr}(A_{\bar{z}}A_{\bar{w}} )\\&+ \int_{\{0\}\times \mathbb{C}^2} dz \wedge dw \wedge d\bar{z} \wedge d\bar{w}\textrm{Tr}(A_{\bar{z}}A_{\bar{w}}).
\fe

The boundary value of $A_{\bar{w}}$ at $t=0$ and $A_{\bar{z}}$ at $t=1$ are denoted as $(A_{\bar{w}})_{\textrm{in}}$ and $(A_{\bar{z}})_{\textrm{out}}$, respectively (with $t$ interpreted as an Euclidean time direction ``in" and ``out" shall denote $t=0$ and $t=1$, respectively). 
We shall first consider the case without non-commutativity, and subsequently turn on non-commutativity.

The 5d CS theory is extended in the BV-BFV formalism to include ghosts, denoted $c$, and antifields. The action\footnote{Note that we shall include an overall factor of $\frac{1}{2}$ for the Chern-Simons action in this section.} 
\begin{eqnarray} 
\frac{1}{2}\int_{I\times \mathbb{C}^2} dz \wedge dw \wedge  \textrm{Tr} \left(\cA \wedge d \cA + \frac{2}{3} \cA \wedge \cA \wedge \cA  \right).
\end{eqnarray} 
is defined in terms of the superfield 
\begin{equation}
\mathcal{A}=c+A+A^*+c^*,
\end{equation}
where $\phi^*$ denotes the BV antifield of a field $\phi$.

To be precise, the ghost field $c$
has ghost number $1$, while the gauge field  
\ie  A=A_t\text{d}t + A_{\bar{w}} \text{d}\bar{w} + A_{\bar{z}} \text{d}\bar{z} \,\, 
\fe 
has ghost number $0$.
Moreover, the two form antifield of the gauge field, \ie A^*=A^{*}_{t\bar{w}} \text{d}t \wedge \text{d}\bar{w} + A^{*}_{t\bar{z}} \text{d}t \wedge \text{d} \bar{z} + A^*_{\bar{w}\bar{z}}d\bar{w}\wedge d \bar{z} \,\,
\fe 
has ghost number $-1$,
while the three-form antifield of the ghost field 
\ie c^*=c^{*}_{t \bar{w} \bar{z}} \text{d}t\wedge \text{d}\bar{w}\wedge \text{d}\bar{z} \,\, 
\fe 
has ghost number $-2$. In addition, we shall decompose $\mathcal{A}$ into its components along $\mathbb{C}^2 $ and the interval, $I$, as 
\ie 
\mathcal{A}=\mathrm{A}+dt\cdot \mathrm{A}_I.
\fe

We further split each field $\phi$ into a residual and quantum fluctuation part as $\phi_{\textrm{res}}+\phi_{\textrm{fl}}$. In what follows we shall use the superscript $1$ to denote the $\bar{z}$ component and $2$ to denote the $\bar{w}$ component of the 1-form gauge field. We also use the notations
$a^1=\mathrm{A}_{\mathrm{fl}}^1, a^2=\mathrm{A}_{\mathrm{fl}}^2, \sigma=\mathrm{A}_{I \text { res }}$. The computation that follows, where we shall evaluate the effective action of 5d CS in the BV-BFV formalism, is analogous to the computation for 3d CS in \cite{Cattaneo:2020lle}. We shall generalize the computation to include matter fields in the second part of this section. 

We also need to specify the gauge-fixing conditions, which amount to specifying a Lagrangian submanifold  of the field space. The field space is defined as 
\ie 
\mathcal{F}=\Omega^{\bullet}\left(I, \Omega_{\bar{\partial}^1}^{0,1} \oplus \Omega_{\bar{\partial}^2}^{0,1} \oplus \Omega^0[1] \oplus \Omega^2[-1]\right),
\fe 
where $\Omega_{\bar{\partial}^1}^{0,1}$  and $\Omega_{\bar{\partial}^2}^{0,1}$ respectively denote anti-holomorphic differential one-forms on $\C_z$ and $\C_w$, respectively, while $\Omega^0$ and $\Omega^2$ denote zero-forms and anti-holomorphic two-forms on $\C_z \times \C_w$.
The field space is fibered over
\ie 
\mathcal{B}=\left(\Omega_{\bar{\partial}^2}^{0,1} \oplus \Omega^0[1]\right) \bigoplus\left(\Omega_{\bar{\partial}^1}^{0,1} \oplus \Omega^0[-1]\right) \quad \ni\left(\left(\mathrm{A}_{\text {in }}^{2}, c_{\text {in }}\right),\left(\mathrm{A}_{\text {out }}^{1}, c_{\text {out }}\right)\right)
\fe
with fiber
\ie
\mathcal{Y}=\Omega^{\bullet}\left(I,\{0\} ; \Omega_{\bar{\partial}^2}^{0,1}\right) \oplus \Omega^{\bullet}\left(I,\{1\} ; \Omega_{\bar{\partial}^1}^{0,1}\right) \oplus \Omega^{\bullet}\left(I,\{0,1\} ; \Omega^0[1]\right) \oplus \Omega^{\bullet}\left(I ; \Omega^2[-1]\right)
\fe
The residual fields form a space defined by the relative cohomology in the $I$-direction:
\ie 
\mathcal{V}=H^{\bullet}\left(I,\{0,1\} ; \Omega^0[1]\right) \oplus H^{\bullet}\left(I ; \Omega^2[-1]\right) \quad \ni\left(d t \cdot \sigma, \mathrm{A}_{\mathrm{res}}^*\right).
\fe 
On this field space we can define the odd symplectic form 
\ie 
\omega_{I\times \C^2 }(\cA, \cA')=\int_{I\times \C^2} dz\wedge dw\wedge \textrm{Tr }\delta \cA \wedge \delta A',
\fe
while the space of boundary fields admits the even symplectic form 
\ie 
\omega^{\partial}_{\partial I\times \C^2 }(\cA, \cA')=\int_{\partial I\times \C^2}dz\wedge dw\wedge \textrm{Tr } \delta \cA \wedge \delta A'.
\fe

The gauge-fixing Lagrangian submanifold in the fiber of $\mathcal{Y} \rightarrow\mathcal{V}$ is determined by setting to zero the relatively exact one-form components of fields along the $I$ direction. The gauge-fixing conditions can thus be specified as 
\ie 
\begin{aligned}
\operatorname{gh}=0: & A=\widetilde{\mathrm{A}}_{\text {out }}^{1}+\widetilde{\mathrm{A}}_{\mathrm{in}}^{2}+a^{1}+a^{2}-d t \cdot \sigma, \\
\mathrm{gh}=1: & c=\widetilde{c}_{\text {out }}+\widetilde{c}_{\text {in }}+c_{\mathrm{fl}}, \\
\mathrm{gh}=-1: & A^*=\mathrm{A}_{\text {res }}^*+\mathrm{A}_{\mathrm{fl}}^* \\
\mathrm{gh}=-2: & c^*=0.
\end{aligned}
\fe
where tilde denotes the discontinuous extension by zero from $t=1$ or $t=0$. This is the \textit{axial gauge}, where fluctuation fields that are both 1-forms along $I$ and forms of any degree along $\C^2$ are set to zero. 
The fluctuation fields further satisfy
\ie 
\left.a^{1}\right|_{t=1}=0,\left.\quad a^{2}\right|_{t=0}=0,\left.\quad c_{\mathrm{fl}}\right|_{t=0}=\left.c_{\mathrm{fl}}\right|_{t=1}=0, \quad \int_0^1 d t \mathrm{~A}_{\mathrm{fl}}^*=0.
\fe

The gauge-fixed kinetic term of the 5d CS action, with relevant polarization boundary terms included, can then be written as 
\ie 
\begin{aligned}
 S =& \frac{1}{2} \int_{I\times \mathbb{C}^2} dz \wedge dw \wedge {\mathrm{Tr}}\left(\cA \wedge d \cA  \right) +\int_{\{1\} \times \C^2 } dz \wedge dw \wedge \textrm{Tr}(\mathrm{A}^1 \wedge \mathrm{A}^2+ c\mathrm{A}^* )\\&-\int_{\{0\} \times \C^2} dz \wedge dw \wedge \textrm{Tr}(\mathrm{A}^2 \wedge \mathrm{A}^1+ c\mathrm{A}^* )\\
=& \int_{I \times \mathbb{C}^2} dz \wedge dw \wedge {\mathrm{Tr}} \left(a^1 \wedge d_I a^2\right)-\int_{I \times \mathbb{C^2}} dz \wedge dw \wedge d t \wedge {\mathrm{Tr}}\left(a^1+a^2\right) \wedge \bar{\partial}_{\mathbb{C}^2} \sigma\\&+\left.\int_{\mathbb{C}^2}  dz \wedge dw \wedge {\mathrm{Tr}} \left(\mathrm{A}_{\text {out }}^1 \wedge a^2\right)\right|_{t=1}-\left.\int_{\mathbb{C}^2} dz \wedge dw \wedge {\mathrm{Tr}} \left(\mathrm{A}_{\text {in }}^2 \wedge a^1\right)\right|_{t=0}  \\&+\int_{I \times \mathbb{C}^2} dz \wedge dw \wedge {\mathrm{Tr}} \left(\mathrm{A}_{\mathrm{fl}}^* \wedge d_I c_{\mathrm{fl}}\right)\\&-\int_{\mathbb{C}^2} dz \wedge dw \wedge {\mathrm{Tr}} \left(\mathrm{A}_{\mathrm{res}}^*+\left.\mathrm{A}_{\mathrm{fl}}^*\right|_{t=1}\right) c_{\mathrm{out}}+\int_{\mathbb{C}^2} dz \wedge dw \wedge {\mathrm{Tr}}\left(\mathrm{A}_{\mathrm{res}}^*+\left.\mathrm{A}_{\mathrm{fl}}^*\right|_{t=0}\right) c_{\mathrm{in}} , 
\end{aligned}
\fe 
where $d_I$ and $\bar{\partial}_{\mathbb{C}^2}$ are the exterior and antiholomorphic derivatives along $I$ and $\mathbb{C}^2 $, respectively, and ``in" and ``out" denote boundary values for fields at $t=0$ and $t=1$ respectively.

The free-field propagators are given by:
\ie 
\left\langle a_b^2(t, z,w)a_c^1\left(t^{\prime}, z^{\prime},w^{\prime}\right)\right\rangle= \hbar \theta\left(t-t^{\prime}\right) \delta_{bc}\delta^{(2)}\left(z-z^{\prime}\right) \delta^{(2)}(w-w') d \bar{z}' d \bar{w}  ,
\fe
\ie 
\left\langle c_{\mathrm{fl},b}(t, z,w) \mathrm{A}_{\mathrm{fl},c}^*\left(t^{\prime}, z^{\prime},w^{\prime}\right)\right\rangle= \hbar\left(\theta\left(t-t^{\prime}\right)-t\right) \delta_{bc}\delta^{(2)}\left(z-z^{\prime}\right) \delta^{(2)}(w-w') d \bar{w}' d \bar{z}'   ,
\fe
where $b$ and $c$ are Lie algebra indices, and where the step function is defined as 
\ie 
\theta(x):= \begin{cases}1, & x > 0 \\ 0, & x<0.\end{cases}
\fe 

Restricting the interaction term of the nonabelian theory to the relevant gauge-fixing Lagrangian submanifold of the space of fields gives us 
\ie \label{ncc1}
S_{\mathrm{int}}=&\frac{1}{6} \int_{I\times \C^2 } dz\wedge dw\langle\mathcal{A},[\mathcal{A}, \mathcal{A}]\rangle\\ =&-\int_{\C^2}dz\wedge dw \wedge \int_I d t\left\langle a^1, \wedge \operatorname{ad}_\sigma a^2\right\rangle +\int_{\C^2} dz\wedge dw \wedge \int_I d t\left\langle c_{\mathrm{fl}}, \operatorname{ad}_\sigma\left(\mathrm{A}_{\mathrm{res}}^*+\mathrm{A}_{\mathrm{fl}}^*\right)\right\rangle, \fe 
where we have used $\langle , \rangle$ to indicate the Killing form on the gauge group, which we have previously denoted as $\textrm{Tr}$ (we shall use both notations interchangeably in what follows). 
The dressed propagators that take into account these interaction terms are 
\ie \label{ncc2}\begin{aligned} &\left\langle a_b^2(t, z,w) a_c^1\left(t^{\prime}, z^{\prime}, w^{\prime}\right)\right\rangle_{\mathrm{dressed}} \\& = \hbar \theta\left(t-t^{\prime}\right) \delta^{(2)}\left(z-z^{\prime} \right) \delta^{(2)}(w-w')d \bar{w}  d \bar{z}' \left( \sum_{k=0}^{\infty} \int_{t^{\prime}<t_1<\cdots<t_k<t} d t_1 \cdots d t_k\left(-\operatorname{ad}_\sigma\right)^k \right)_{bc}\\ & = \hbar \theta\left(t-t^{\prime}\right) \bigg( e^{-\left(t-t^{\prime}\right) \operatorname{ad}_\sigma}\bigg)_{bc} \delta^{(2)}\left(z-z^{\prime}\right) \delta^{(2)}(w-w') d \bar{z}' d \bar{w}  ,\end{aligned}\fe 
and
\ie \label{ncc3}\begin{aligned} &\left\langle c_{\mathrm{fl},b}(t, z,w ) \mathrm{A}_{\mathrm{fl},c}^*\left(t^{\prime}, z^{\prime},w^{\prime}\right)\right\rangle_{\mathrm{dressed}}=\hbar \delta^{(2)}\left(z-z^{\prime}\right) \delta^{(2)}(w-w')  d \bar{w}' d \bar{z}' . \\ & \cdot \left( \sum_{k=0}^{\infty} \int_{t_1, \ldots, t_k \in[0,1]} d t_1 \cdots d t_k\left(\theta\left(t-t_1\right)-t\right)\left(\theta\left(t_1-t_2\right)-t_1\right) \cdots\left(\theta\left(t_k-t^{\prime}\right)-t_k\right)\left(-\mathrm{ad}_\sigma\right)^k\right)_{bc}
\\=&\hbar \delta^{(2)}(z-z')\delta^{(2)}(w-w')d\bar{w}' d\bar{z}' \left(\frac{e^{\left(t^{\prime}-t+\theta\left(t-t^{\prime}\right)\right) \operatorname{ad}_\sigma}-e^{t^{\prime} \mathrm{ad}_\sigma}}{e^{\operatorname{ad}_\sigma}-1}\right)_{bc}.
\end{aligned}\fe

As detailed in Appendix \ref{B}, the effective action can be computed with these propagators using the techniques of \cite{Cattaneo:2020lle}, by computing the relevant Feynman diagrams shown in Figure \ref{feyn}, 
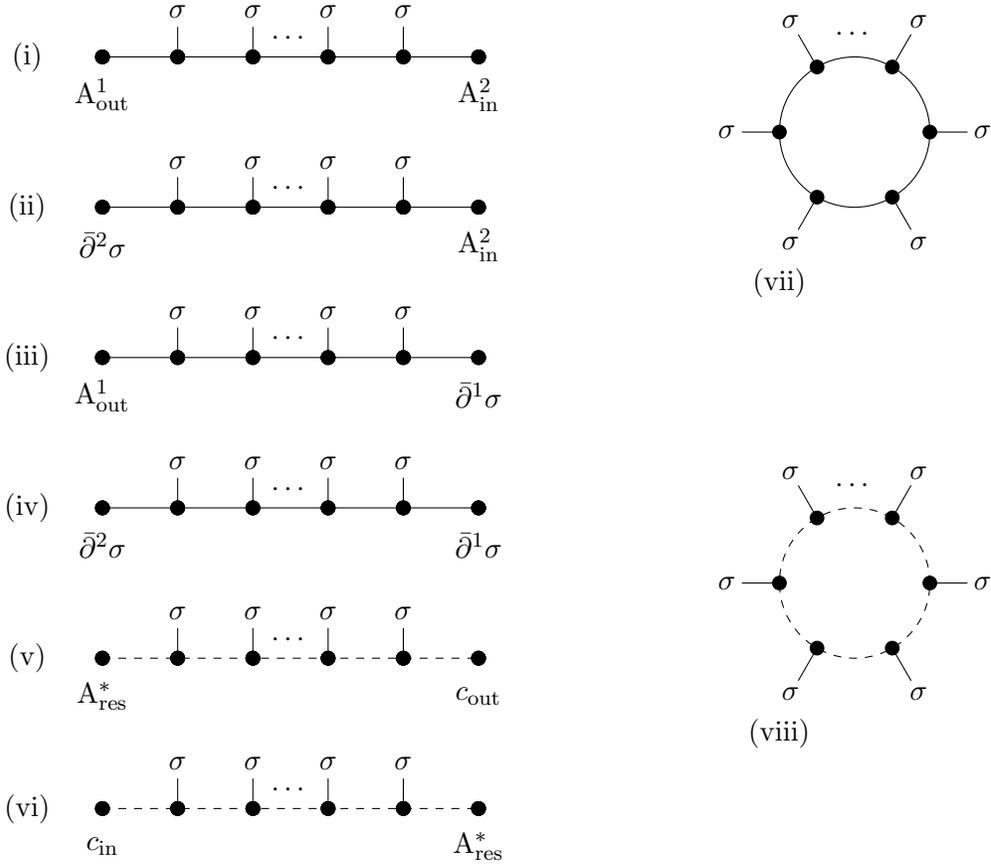
\begin{figure}

\begin{tikzpicture}
\tikzset{
    physical/.style={decorate, draw=black, thin}, 
    ghost/.style={draw=black, thin}, 
    antifield/.style={double, draw=black, thin}, 
    vertex/.style={fill=black, draw=black, shape=circle, radius=0.5pt, scale=0.5}, 
    interaction/.style={circle, draw=black, fill=black, radius=0.25pt, scale=0.5} 
}
\begin{scope}[shift={(-7,7)}]
\draw (0,0) -- (5,0);

\foreach \x in {1, 2, 3, 4}
{
    \draw (\x, 0) -- (\x, 0.4);
    \node at (\x, 0.6) {$\sigma$};
    \node[interaction]at (0,0) {};
    \node[interaction]at (1,0) {};
    \node[interaction]at (2,0) {};
    \node[interaction]at (3,0) {};
    \node[interaction]at (4,0) {};
    \node[interaction]at (5,0) {};

}

\node at (0, -0.5) {$  \mathrm{A}^{1}_{\text{out}}$};
\node at (5, -0.5) {$\mathrm{A}^{2}_{\text{in}}  $};

\node at (2.5, 0.25) {$\cdots$};
    \node at (-1,0) {(i)};
\end{scope}

\begin{scope}[shift={(-7,5)}]
\draw (0,0) -- (5,0);

\foreach \x in {1, 2, 3, 4}
{
    \draw (\x, 0) -- (\x, 0.4);
    \node at (\x, 0.6) {$\sigma$};
    \node[interaction]at (0,0) {};
    \node[interaction]at (1,0) {};
    \node[interaction]at (2,0) {};
    \node[interaction]at (3,0) {};
    \node[interaction]at (4,0) {};
    \node[interaction]at (5,0) {};

}

\node at (0, -0.5) {$  \bar{\partial}^2 \sigma$};
\node at (5, -0.5) {$\mathrm{A}^{2}_{\text{in}}  $};

\node at (2.5, 0.25) {$\cdots$};
    
    \node at (-1,0) {(ii)};
\end{scope}

\begin{scope}[shift={(-7,3)}]
\draw (0,0) -- (5,0);

\foreach \x in {1, 2, 3, 4}
{
    \draw (\x, 0) -- (\x, 0.4);
    \node at (\x, 0.6) {$\sigma$};
    \node[interaction]at (0,0) {};
    \node[interaction]at (1,0) {};
    \node[interaction]at (2,0) {};
    \node[interaction]at (3,0) {};
    \node[interaction]at (4,0) {};
    \node[interaction]at (5,0) {};

}

\node at (0, -0.5) {$  \mathrm{A}^{1}_{\text{out}}$};
\node at (5, -0.5) {$\bar{\partial}^1{\sigma}  $};

\node at (2.5, 0.25) {$\cdots$};
    
    \node at (-1,0) {(iii)};
\end{scope}

\begin{scope}[shift={(-7,1)}]
\draw (0,0) -- (5,0);

\foreach \x in {1, 2, 3, 4}
{
    \draw (\x, 0) -- (\x, 0.4);
    \node at (\x, 0.6) {$\sigma$};
    \node[interaction]at (0,0) {};
    \node[interaction]at (1,0) {};
    \node[interaction]at (2,0) {};
    \node[interaction]at (3,0) {};
    \node[interaction]at (4,0) {};
    \node[interaction]at (5,0) {};

}

\node at (0, -0.5) {$  \bar{\partial}^2\sigma$};
\node at (5, -0.5) {$\bar{{\partial}}^1\sigma  $};

\node at (2.5, 0.25) {$\cdots$};
    \node at (-1,0) {(iv)};
\end{scope}

\begin{scope}[shift={(-7,-1)}]
\draw [dashed](0,0) -- (5,0);

\foreach \x in {1, 2, 3, 4}
{
    \draw (\x, 0) -- (\x, 0.4);
    \node at (\x, 0.6) {$\sigma$};
    \node[interaction]at (0,0) {};
    \node[interaction]at (1,0) {};
    \node[interaction]at (2,0) {};
    \node[interaction]at (3,0) {};
    \node[interaction]at (4,0) {};
    \node[interaction]at (5,0) {};

}

\node at (0, -0.5) {$ \mathrm{A}^{*}_{\text{res}}$};
\node at (5, -0.5) {$c_{\text{out}}  $};

\node at (2.5, 0.25) {$\cdots$};
    
    \node at (-1,0) {(v)};
\end{scope}

\begin{scope}[shift={(-7,-3)}]
\draw [dashed] (0,0) -- (5,0);

\foreach \x in {1, 2, 3, 4}
{
    \draw (\x, 0) -- (\x, 0.4);
    \node at (\x, 0.6) {$\sigma$};
    \node[interaction]at (0,0) {};
    \node[interaction]at (1,0) {};
    \node[interaction]at (2,0) {};
    \node[interaction]at (3,0) {};
    \node[interaction]at (4,0) {};
    \node[interaction]at (5,0) {};

}

\node at (0, -0.5) {$  c_{\text{in}}$};
\node at (5, -0.5) {$\mathrm{A}^{*}_{\text{res}}  $};

\node at (2.5, 0.25) {$\cdots$};
    \node at (-1,0) {(vi)};
\end{scope}

\begin{scope}[shift={(3,6)}]
    \draw[physical] (0,0) circle (1cm);
    
    \foreach \angle in {0, 60, 120, 180,240,300} {
        \draw[physical] (\angle:1cm) -- ++(\angle:0.5cm);
    \node at (\angle:1.7cm) {$\sigma$};
        \node[interaction] at (\angle:1cm) {}; 
         
    }
\node at (90:1.3cm) {$\cdots$};
    \node at (-1,-2) {(vii)};
\end{scope}
\begin{scope}[shift={(3,0)}]
    \draw[dashed] (0,0) circle (1cm);
    
    \foreach \angle in {0, 60, 120, 180,240,300} {
        \draw[physical] (\angle:1cm) -- ++(\angle:0.5cm);
            
    \node at (\angle:1.7cm) {$\sigma$};
        \node[interaction] at (\angle:1cm) {}; 
    }
\node at (90:1.3cm) {$\cdots$};
    \node at (-1,-2) {(viii)};
\end{scope}

\end{tikzpicture}


\caption{Feynman diagrams}
\label{feyn}
\end{figure}
 giving us a gauged 4d chiral WZW model of the form 

\begin{equation}
\begin{aligned}
\label{efact}
S^{\mathrm{eff}}=&\int_{\C^2}dz \wedge dw\left(\left\langle\mathrm{A}_{\text {out }}^1, g \mathrm{~A}_{\mathrm{in}}^2 g^{-1}\right\rangle + \left\langle\mathrm{A}_{\text {out }}^1, \bar{\partial}^2 g \cdot g^{-1}\right\rangle + \left\langle\mathrm{A}_{\text {in }}^2, g^{-1} \bar{\partial}^1 g\right\rangle\right. \\ & \left.\quad-\left\langle\mathrm{A}_{\text {res }}^*, F_{+}\left(\operatorname{ad}_{-\log g}\right)  c_{\text {out }}+F_{-}\left(\operatorname{ad}_{-\log g}\right) c_{\text {in }}\right\rangle\right)+\mathrm{WZW}(g)-i \hbar \mathbb{W}
\end{aligned}
\end{equation}
where $\mathrm{WZW}(g)$ is defined in Appendix \ref{B}, with $g$ related to $\sigma$ via $g=e^{-\sigma}$, where we have denoted the boundary values of the fields at $0$ and $1$ on $I=[0,1]$ with subscripts $``\textrm{in}"$ and $``\textrm{out}"$, where
\begin{equation}\label{fpm}
F_{+}(x)=\frac{x}{1-e^{-x}}, \textrm{   }F_{-}(x)=-\frac{x}{e^{x}-1},
\end{equation}
and where $i \hbar \mathbb{W}$ is a divergent term that can be interpreted as a change of path integral measure \cite{Cattaneo:2020lle}.

We note that, at least for abelian gauge group, in analogy to the 3d-2d correspondence, the effective action we have derived is in fact a generating functional for the current algebra studied in the previous section, i.e., appropriate functional derivatives of the generating functional furnishes the quantum current algebra in OPE form.

We have not considered non-commutativity explicitly in our analysis thus far. However, equations  \eqref{ncc1}, \eqref{ncc2}, and \eqref{ncc3} generalize straightforwardly in the presence of non-commutativity, with the adjoint map $\textrm{ad}_{\sigma}$ now generalized to its non-commutative counterpart. The computation of the non-commutative dual 4d WZW theory then follows from a direct generalization of the computation. A discussion of the generalization of the necessary mathematical formulas to the non-commutative case can be found in Appendix \eqref{ncgen}.

\subsection{BV-BFV Partition Function for 5d Chern-Simons-Matter Theory}

We would now like to extend the BV-BFV quantization of 5d Chern-Simons theory to include matter fields in the adjoint representation. To this end, we consider the action given by:
\begin{equation}
S_{\text{bulk,matter}} = \int_{\mathbb{R} \times \mathbb{C}^2} dz \wedge dw \wedge {\mathrm{Tr}}(\boldsymbol{\eta} \wedge d_{\mathcal{A}}\boldsymbol{\phi}),
\end{equation}
where $\boldsymbol{\eta}$ and $\boldsymbol{\phi}$ are BV superfields that include the physical matter fields as components.
Additionally, we consider the boundary terms:
\begin{equation}
S_{\text{boundary, matter}} =  \int_{\{0\}\times \mathbb{C}^2 } dz \wedge dw \wedge {\mathrm{Tr}}( \boldsymbol{\eta} \wedge  \boldsymbol{\phi}) + \int_{\{1\}\times \mathbb{C}^2 } dz \wedge dw \wedge {\mathrm{Tr}}( \boldsymbol{\eta} \wedge  \boldsymbol{\phi}),
\end{equation}
such that the total matter action is $S_{\textrm{matter}}=S_{\textrm{bulk,matter}}+S_{\textrm{boundary,matter}}$.

The superfields $\boldsymbol{\eta}$ and $\boldsymbol{\phi}$ contain the following components:
\ie
\boldsymbol{\eta} &= {\eta} + {\phi}^* , \\
\boldsymbol{\phi} &= {\phi}+ {\eta}^* ,
\fe
where:
\begin{itemize}
   
    \item ${\eta}$ and ${\phi}$ are matter fields,
    \item ${\eta}^*$ and $\phi^*$ are antifields associated with ${\eta}$ and ${\phi}$, respectively.
    
\end{itemize}

The BV-BFV field space is now extended to include the matter fields $\boldsymbol{\eta}$ and $\boldsymbol{\phi}$ along with the gauge fields. 
The odd symplectic form 
associated with the matter fields is 
\ie 
\omega_{I\times \C^2 }(\boldsymbol{\eta}, \boldsymbol{\phi})=\int_{I\times \C^2} \textrm{Tr }\delta \boldsymbol{\eta} \wedge \delta \boldsymbol{\phi},
\fe
while the space of boundary matter fields admits the even symplectic form 
\ie 
\omega^{\partial}_{\partial I\times \C^2 }(\boldsymbol{\eta}, \boldsymbol{\phi})=\int_{\partial I\times \C^2} \textrm{Tr } \delta \boldsymbol{\eta} \wedge \delta \boldsymbol{\phi}.
\fe
The gauge-fixing conditions for the matter fields are similar to those used for the gauge fields, focusing on setting relatively exact components to zero:
\ie
\begin{aligned}
\text{gh}=0: & \quad \boldsymbol{\phi} = \widetilde{\phi}_{\text{out}}^1 + \widetilde{\phi}_{\text{in}}^2 + a^{1}_{\phi}+a^{2}_{\phi} - d t \cdot \sigma_{{\phi}}, \quad \boldsymbol{\eta} = \widetilde{{\eta}}_{\text{out}}^1 + \widetilde{{\eta}}_{\text{in}}^2 + a^{1}_{{\eta}}+a^{2}_{{\eta}} - d t \cdot \sigma_{{\eta}},\\
\mathrm{gh}=-1: & \quad \boldsymbol{\phi}^*=\mathrm{\phi}_{\text {res }}^*+\mathrm{\phi}_{\mathrm{fl}}^*, \quad  \quad {\boldsymbol{\eta}}^*=\mathrm{{\eta}}_{\text {res }}^*+\mathrm{{\eta}}_{\mathrm{fl}}^*.
\end{aligned}
\fe
Moreover, the matter fluctuation fields satisfy
\ie 
\left.a_{\eta}^{1}\right|_{t=1}=0,\quad\left.a_{\phi}^{1}\right|_{t=1}=0,\left.\quad a_{\eta}^{2}\right|_{t=0}=0,\left.\quad  a_{\phi}^{2}\right|_{t=0}=0, \quad \int_0^1 d t\textrm{ } \eta_{\mathrm{fl}}^*=0, \quad \int_0^1 d t \textrm{ } \phi_{\mathrm{fl}}^*=0.
\fe

Now, the kinetic and boundary terms of the gauge-fixed matter action take the form:
\begin{equation}
\begin{aligned}
&S_{\textrm{kin}}+S_{\textrm{boundary,matter}}\\=&\int_{I\times \mathbb{C}^2} dz\wedge dw \wedge \textrm{Tr} (a^1_{\eta}d_I a_{\phi}^2 + a^2_{\eta}d_I a_{\phi}^1 - (a^1_{\eta }+ a^2_{\eta}) d_{\C\times \C}\sigma_{\phi} - \sigma_{\eta}d_{\C\times \C} (a^1_{\phi }+ a^2_{\phi}))\\&+\int_{\{0\} \times \C^2}dz\wedge dw \wedge \textrm{Tr} (\tilde{\eta}_{\textrm{in}}^2\wedge a^1_{\phi})+\int_{\{1\} \times \C^2}dz\wedge dw \wedge \textrm{Tr} ( a^2_{\eta} \wedge \tilde{\phi}_{\textrm{out}}^1)
\end{aligned}
\end{equation}
where $d_I$ and $\bar{\partial}_{\mathbb{C} \times \mathbb{C}}$ are the exterior and antiholomorphic derivatives along $I$ and $\mathbb{C}^2 $, respectively.
The interaction term for the matter fields in the non-abelian theory can be written as
\ie
S_{\mathit{int}} =-\int_{I \times \C^2}dz \wedge dw \wedge dt \wedge \textrm{Tr} \bigg(& -\textrm{ad}_{\sigma_{\phi}}(\phi^*_{\textrm{res}}+\phi^*_{\textrm{fl}})c_{\textrm{fl}}  +\textrm{ad}_{\sigma_{\eta}}(\eta^*_{\textrm{res}}+\eta^*_{\textrm{fl}})c_{\textrm{fl}} -a^1_{\eta} \textrm{ad}_{\sigma}a^2_{\phi}-a^2_{\eta} \textrm{ad}_{\sigma}a^1_{\phi} \\&+(\textrm{ad}_{\sigma_{\eta}} a^1) a^2_{\phi} +(\textrm{ad}_{\sigma_{\eta}} a^2) a^1_{\phi}-a^2_{\eta }\textrm{ad}_{\sigma_{\phi}}a^1 -a^1_{\eta }\textrm{ad}_{\sigma_{\phi}}a^2 \bigg) . 
\fe

The dressed  propagators for the matter fields are given by:
\begin{align}\label{pq3}
\left\langle a^2_{{\eta},b}(t, z, w)  a^1_{{\phi},c}(t', z', w') \right\rangle_{\mathrm{dressed}} 
&= \hbar \theta(t - t') \delta^{(2)}(z - z') \delta^{(2)}(w - w')  d\bar{z}' d\bar{w} \nonumber\\
&\quad \times \left( \sum_{k=0}^{\infty} \int_{t' < t_1 < \cdots < t_k < t} dt_1 \cdots dt_k \left(-\operatorname{ad}_{-\log g}\right)^k \right)_{bc}\nonumber \\
&= \hbar \theta(t - t') \left (e^{-(t - t') \operatorname{ad}_{-\log g}} \right)_{bc}\delta^{(2)}(z - z') \delta^{(2)}(w - w')  d\bar{z}' d\bar{w},
\end{align}
and 
\begin{align}
\left\langle a^2_{{\phi},b}(t, z, w)  a^1_{{\eta},c}(t', z', w') \right\rangle_{\mathrm{dressed}} 
&= \hbar \theta(t - t') \delta^{(2)}(z - z') \delta^{(2)}(w - w')  d\bar{z}' d\bar{w} \nonumber\\
&\quad \times \left( \sum_{k=0}^{\infty} \int_{t' < t_1 < \cdots < t_k < t} dt_1 \cdots dt_k \left(-\operatorname{ad}_{-\log g}\right)^k \right)_{bc}\nonumber \\
&= \hbar \theta(t - t') \left(e^{-(t - t') \operatorname{ad}_{-\log g}} \right)_{bc}\delta^{(2)}(z - z') \delta^{(2)}(w - w')  d\bar{z}' d\bar{w}.
\end{align}

As shown in Appendix \ref{4dwzm}, the effective 4d action for the matter fields can be computed from the Feynman diagrams in Figure~\ref{feyn} to be:
\ie
\begin{aligned}
 S^{\mathrm{eff}} =&\int_{\C^2}dz\wedge dw \wedge  \Bigg\{ \langle \tilde{\phi}_{\text{out}}^1, e^{-\operatorname{ad}_{-\log g}}  \tilde{\eta}_{\text{in}}^2 \rangle +\left\langle{\tilde{\phi}}_{\mathrm{out}}^1, \frac{1-e^{-\mathrm{ad}_{-\log g}}}{\mathrm{ad}_{-\log g}}  {\partial}^2 \sigma_{{\eta}}\right\rangle \\&-\left\langle {\partial}^1 \sigma_{{\phi}}, \frac{1-e^{-\mathrm{ad}_{-\log g}}}{\mathrm{ad}_{-\log g}} \tilde{\eta}_{\mathrm{in}}^2 \right\rangle 
 -\left\langle \partial^1  \sigma_{{\phi}}, \frac{e^{-\mathrm{ad}_{-\log g}}+\mathrm{ad}_{-\log g}-1}{\left(\operatorname{ad}_{-\log g}\right)^2}  \partial^2\sigma_{{\eta}}\right\rangle \\& -\left\langle \partial^1 \sigma_{{\eta}}, \frac{e^{-\mathrm{ad}_{-\log g}}+\mathrm{ad}_{-\log g}-1}{\left(\operatorname{ad}_{-\log g}\right)^2}  \partial^2\sigma_{{\phi}}\right\rangle \\&-\left\langle (\textrm{ad}_{\sigma_{\phi}}{\phi}_{\text{res}}^*  -\textrm{ad}_{\sigma_{\eta}}{\eta}_{\text{res}}^*),  \left( \frac{1}{\textrm{ad}_{-\log g}} - \frac{1}{1-e^{-\textrm{ad}_{-\log g}}}\right) c_{\text{out}} \right\rangle \\& +\left\langle (\textrm{ad}_{\sigma_{\phi}}{\phi}_{\text{res}}^*  -\textrm{ad}_{{\sigma}_{\eta}}{\eta}_{\text{res}}^*),  \left( \frac{1}{\textrm{ad}_{-\log g}} - \frac{1}{e^{\textrm{ad}_{-\log g}}-1}\right) c_{\text{in}} \right\rangle
 \Bigg\}.\end{aligned}\fe
where, as before, the boundary values of the fields at $0$ and $1$ on $I=[0,1]$ are denoted with subscripts $``\textrm{in}"$ and $``\textrm{out}"$, and $\sigma = -\textrm{log }g$. 
The effective matter action involves  nontrivial interactions between matter fields and the WZW field $g$. In particular, one finds terms that resemble kinetic terms for $\sigma_{\phi}$ and $\sigma_{\eta}$.

The complete boundary 4d WZW-matter effective action is therefore
\ie
\begin{aligned}
 &S^{\mathrm{eff}}\\ =&\mathrm{WZW}(g) + \int_{\C^2}dz\wedge dw \wedge \Bigg\{\langle\mathrm{A}_{\text {out }}^1, g \mathrm{~A}_{\text {in }}^2 g^{-1} \rangle +\left\langle\mathrm{A}_{\text {out }}^1,{\bar{\partial}^2} g \cdot g^{-1}\right\rangle +\left\langle\mathrm{A}_{\text {in }}^2, g^{-1} \bar{\partial}^1 g\right\rangle \\
& -\left\langle\mathrm{A}_{\text {res }}^*, F_{+}\left(\operatorname{ad}_{-\log g}\right) \cdot c_{\text {out }}+F_{-}\left(\operatorname{ad}_{-\log g}\right) \cdot c_{\text {in }}\right\rangle + \left\langle   \tilde{\phi}_{\mathrm{out}}^1, e^{-\mathrm{ad}_{-\log g}} \tilde{\eta}_{\text {in }}^2\right\rangle \\& +\left\langle \tilde{\phi}_{\mathrm{out}}^1, \frac{1-e^{-\mathrm{ad}_{-\log g}}}{\mathrm{ad}_{-\log g}}  {\partial}^2 \sigma_{{\eta}}\right\rangle -\left\langle {\partial}^1 \sigma_{{\phi}}, \frac{1-e^{-\mathrm{ad}_{-\log g}}}{\mathrm{ad}_{-\log g}} \tilde{\eta}_{\mathrm{in}}^2 \right\rangle 
 \\&  -\left\langle \partial^1  \sigma_{{\phi}}, \frac{e^{-\mathrm{ad}_{-\log g}}+\mathrm{ad}_{-\log g}-1}{\left(\operatorname{ad}_{-\log g}\right)^2}  \partial^2\sigma_{{\eta}}\right\rangle -\left\langle \partial^1 \sigma_{{\eta}}, \frac{e^{-\mathrm{ad}_{-\log g}}+\mathrm{ad}_{-\log g}-1}{\left(\operatorname{ad}_{-\log g}\right)^2}  \partial^2\sigma_{{\phi}}\right\rangle \\&-\left\langle (\textrm{ad}_{\sigma_{\phi}}{\phi}_{\text{res}}^*  -\textrm{ad}_{\sigma_{\eta}}{\eta}_{\text{res}}^*),  \left( \frac{1}{\textrm{ad}_{-\log g}} - \frac{1}{1-e^{-\textrm{ad}_{-\log g}}}\right) c_{\text{out}} \right\rangle  \\& +\left\langle (\textrm{ad}_{\sigma_{\phi}}{\phi}_{\text{res}}^*  -\textrm{ad}_{\sigma_{\eta}}{\eta}_{\text{res}}^*),  \left( \frac{1}{\textrm{ad}_{-\log g}} - \frac{1}{e^{\textrm{ad}_{-\log g}}-1}\right) c_{\text{in}} \right\rangle
-i \hbar \mathbb{W} \Bigg\}.\end{aligned}\fe

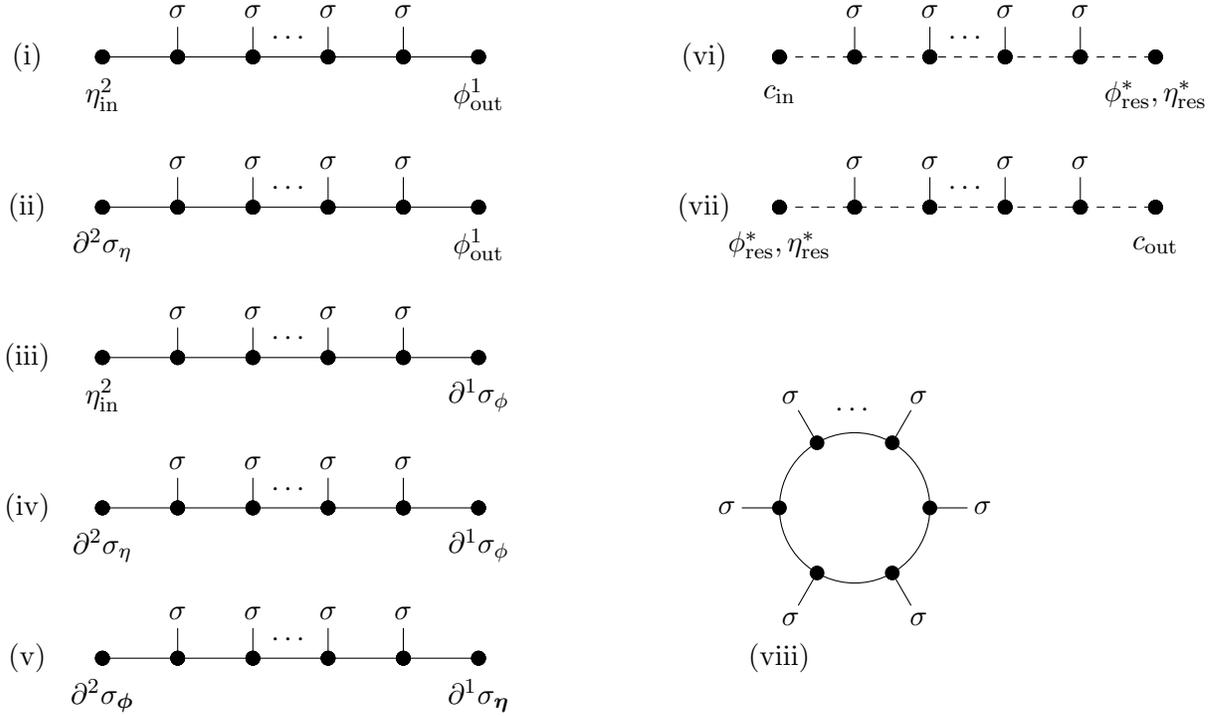
\begin{figure}

\begin{tikzpicture}
\tikzset{
    physical/.style={decorate, draw=black, thin}, 
    ghost/.style={draw=black, thin}, 
    antifield/.style={double, draw=black, thin}, 
    vertex/.style={fill=black, draw=black, shape=circle, radius=0.5pt, scale=0.5}, 
    interaction/.style={circle, draw=black, fill=black, radius=0.25pt, scale=0.5} 
}
\begin{scope}[shift={(-7,7)}]
\draw (0,0) -- (5,0);

\foreach \x in {1, 2, 3, 4}
{
    \draw (\x, 0) -- (\x, 0.4);
    \node at (\x, 0.6) {$\sigma$};
    \node[interaction]at (0,0) {};
    \node[interaction]at (1,0) {};
    \node[interaction]at (2,0) {};
    \node[interaction]at (3,0) {};
    \node[interaction]at (4,0) {};
    \node[interaction]at (5,0) {};

}

\node at (0, -0.5) {$  {\eta}^2_{\text{in}}$};
\node at (5, -0.5) {${\phi}^{1}_{\text{out}}  $};

\node at (2.5, 0.25) {$\cdots$};
    \node at (-1,0) {(i)};
\end{scope}

\begin{scope}[shift={(-7,5)}]
\draw (0,0) -- (5,0);

\foreach \x in {1, 2, 3, 4}
{
    \draw (\x, 0) -- (\x, 0.4);
    \node at (\x, 0.6) {$\sigma$};
    \node[interaction]at (0,0) {};
    \node[interaction]at (1,0) {};
    \node[interaction]at (2,0) {};
    \node[interaction]at (3,0) {};
    \node[interaction]at (4,0) {};
    \node[interaction]at (5,0) {};

}

\node at (0, -0.5) {$  {\partial}^2{\sigma}_{{\eta}}$};
\node at (5, -0.5) {${\phi}_{\text{out}}^1  $};

\node at (2.5, 0.25) {$\cdots$};
    
    \node at (-1,0) {(ii)};
\end{scope}

\begin{scope}[shift={(-7,3)}]
\draw (0,0) -- (5,0);

\foreach \x in {1, 2, 3, 4}
{
    \draw (\x, 0) -- (\x, 0.4);
    \node at (\x, 0.6) {$\sigma$};
    \node[interaction]at (0,0) {};
    \node[interaction]at (1,0) {};
    \node[interaction]at (2,0) {};
    \node[interaction]at (3,0) {};
    \node[interaction]at (4,0) {};
    \node[interaction]at (5,0) {};

}

\node at (0, -0.5) {$ {\eta}^2_{\text{in}}$};
\node at (5, -0.5) {$\partial^1 \sigma_{{\phi}}  $};

\node at (2.5, 0.25) {$\cdots$};
    
    \node at (-1,0) {(iii)};
\end{scope}

\begin{scope}[shift={(-7,1)}]
\draw (0,0) -- (5,0);

\foreach \x in {1, 2, 3, 4}
{
    \draw (\x, 0) -- (\x, 0.4);
    \node at (\x, 0.6) {$\sigma$};
    \node[interaction]at (0,0) {};
    \node[interaction]at (1,0) {};
    \node[interaction]at (2,0) {};
    \node[interaction]at (3,0) {};
    \node[interaction]at (4,0) {};
    \node[interaction]at (5,0) {};

}

\node at (0, -0.5) {$  \partial^2 \sigma_{{\eta}}  $};
\node at (5, -0.5) {$\partial^1 \sigma_{{\phi}}  $};

\node at (2.5, 0.25) {$\cdots$};
    \node at (-1,0) {(iv)};
\end{scope}

\begin{scope}[shift={(-7,-1)}]
\draw (0,0) -- (5,0);

\foreach \x in {1, 2, 3, 4}
{
    \draw (\x, 0) -- (\x, 0.4);
    \node at (\x, 0.6) {$\sigma$};
    \node[interaction]at (0,0) {};
    \node[interaction]at (1,0) {};
    \node[interaction]at (2,0) {};
    \node[interaction]at (3,0) {};
    \node[interaction]at (4,0) {};
    \node[interaction]at (5,0) {};

}

\node at (0, -0.5) {$  {\partial}^2 \sigma_{\boldsymbol{\phi}}$};
\node at (5, -0.5) {$\partial^1 \sigma_{\boldsymbol{\eta}}  $};

\node at (2.5, 0.25) {$\cdots$};
    
    \node at (-1,0) {(v)};
\end{scope}


\begin{scope}[shift={(2,5)}]
\draw [dashed](0,0) -- (5,0);

\foreach \x in {1, 2, 3, 4}
{
    \draw (\x, 0) -- (\x, 0.4);
    \node at (\x, 0.6) {$\sigma$};
    \node[interaction]at (0,0) {};
    \node[interaction]at (1,0) {};
    \node[interaction]at (2,0) {};
    \node[interaction]at (3,0) {};
    \node[interaction]at (4,0) {};
    \node[interaction]at (5,0) {};

}

\node at (0, -0.5) {$\mathrm{\phi}^{*}_{\text{res}}, \mathrm{\eta}^{*}_{\text{res}} $};
\node at (5, -0.5) {$c_{\text{out}}  $};

\node at (2.5, 0.25) {$\cdots$};
    
    \node at (-1,0) {(vii)};
\end{scope}

\begin{scope}[shift={(2,7)}]
\draw [dashed] (0,0) -- (5,0);

\foreach \x in {1, 2, 3, 4}
{
    \draw (\x, 0) -- (\x, 0.4);
    \node at (\x, 0.6) {$\sigma$};
    \node[interaction]at (0,0) {};
    \node[interaction]at (1,0) {};
    \node[interaction]at (2,0) {};
    \node[interaction]at (3,0) {};
    \node[interaction]at (4,0) {};
    \node[interaction]at (5,0) {};

}

\node at (0, -0.5) {$  c_{\text{in}}$};
\node at (5, -0.5) {$\mathrm{\phi}^{*}_{\text{res}}, \mathrm{\eta}^{*}_{\text{res}} $};

\node at (2.5, 0.25) {$\cdots$};
    \node at (-1,0) {(vi)};
\end{scope}

\begin{scope}[shift={(3,1)}]
    \draw[physical] (0,0) circle (1cm);
    
    \foreach \angle in {0, 60, 120, 180,240,300} {
        \draw[physical] (\angle:1cm) -- ++(\angle:0.5cm);
    \node at (\angle:1.7cm) {$\sigma$};
        \node[interaction] at (\angle:1cm) {}; 
         
    }
\node at (90:1.3cm) {$\cdots$};
    \node at (-1,-2) {(viii)};
\end{scope}

\end{tikzpicture}

\caption{Feynman diagrams for matter fields}
\label{feyn}
\end{figure}

We have thus obtained a 4d WZW-matter theory as the holographic dual of the 5d Chern-Simons-matter theory, by evaluating its partition function in the BV-BFV formalism. We have thus evaluated the partition function for twisted M-theory on a general conifold in terms of a boundary effective action, which governs the holographic dual of M-theory in this background.

In the case where matter is turned off, the expectation is that correlation functions of line operators in the bulk ought to be captured by correlation functions of local operators in the boundary, in analogy with the 3d CS/2d WZW correspondence. In the case with matter, the  holographic dictionary between observables in the bulk and observables in the boundary should be a generalization of the case without matter, and we shall leave the precise identification of this dictionary to future work.

\section{Conclusion}
In this work, we have generalized Costello's description of twisted M-theory defined on a Taub-NUT background in terms of 5d non-commutative Chern-Simons theory. We investigated twisted M-theory on a general conifold, and obtained a description in terms of a 5d non-commutative Chern-Simons-matter theory.

In order to analyze this 5d non-commutative Chern-Simons-matter theory with a boundary, we explicitly established the gauge invariant coupling between the 5d Chern-Simons theory and a gauged 4d chiral WZW model, both in the commutative and non-commutative cases. We then generalized this result to the case where the bulk theory is coupled to matter fields. In this system, it is not possible to simply obtain the WZW model by imposing boundary conditions on the Chern-Simons gauge field. However, we  generalized a formalism used previously in the case of the 3d-2d correspondence \cite{Chu:2009ms, Berman:2009xd} to obtain a gauge invariant coupling to a boundary WZW model without imposing boundary conditions. 

Motivated by this result, and by the 3d-2d correspondence of 3d Chern-Simons theory and the 2d WZW model, we then obtained a new 5d-4d correspondence between 5d non-commutative Chern-Simons theory and the gauged 4d chiral WZW model using the BV-BFV formalism. These results were then generalized to the case with matter, furnishing a holographic description of twisted M-theory on a general conifold. 

There are many possible directions for future research. 
Firstly, it is known that the line operators in 5d Chern-Simons theory are associated with representations of the affine Yangian or its 
variants, depending on the choice of holomorphic surface the 5d theory is defined on. 
Given the mathematically rigorous nature of the BV-BFV formalism, we expect the 5d-4d correspondence we derived to provide important mathematical results. In particular, we expect a Kazhdan-Luzstig type correspondence relating the representations of the affine Yangian and the current algebra derived in this work. Moreover, we expect that this formalism allows us to construct conformal blocks associated with the affine Yangian, that give rise to correlation functions of local operators in the 4d chiral WZW model that can be identified with correlation functions of Wilson lines in the bulk.

Although we have used the BV-BFV technique for the specific goal of obtaining a 4d effective action for the 5d Chern-Simons-matter theory, the computation presented here is a prototype for holographic duality for Chern-Simons-matter theory of any dimension. 
 In particular, it is likely that the techniques utilized in our work can be applied to find a holographic dual of the ABJM model \cite{Aharony:2008ug}, as well as the twisted BLG and ABJM models studied in \cite{Okazaki:2015fiq}. 
It would also be interesting to provide an M-theoretic interpretation of the boundaries that we include in our setup in terms of an M-brane or domain wall in twisted M-theory.

In addition, we note that it has been argued that non-commutative Chern-Simons theory is the proper framework for studying the fractional quantum Hall effect. It has  been explicitly demonstrated that 
the Laughlin electron theory of the fractional
quantum Hall effect can be described by a non-commutative Chern Simons theory \cite{Susskind:2001fb, Hellerman:2001rj}.  It is known that the 2d WZW model can be used to study edge states in quantum Hall systems \cite{Daoud:2006nf}. Thus, the edge excitations of higher dimensional quantum hall systems are expected to be described using higher dimensional chiral gauged WZW models \cite{Karabali:2004km}, including the chiral 4d WZW model we have studied. Moreover, electrons in the fractional quantum Hall effect can be described via
a Chern-Simons matrix model, and the quantum algebra of its observables can be identified in the large $N$ limit with a variant of the affine Yangian known as the deformed double current algebra \cite{Hu:2023eyx, Hu:2024waw}, suggesting a further (generalizable) relation between 5d Chern-Simons theory and the fractional quantum Hall effect. 

  Integrability in the context of 5d  non-commutative  Chern-Simons theory has recently been further  understood, in particular with regard to interpreting Miura operators (that realize $W$-algebras) as M-brane intersections \cite{Haouzi:2024qyo, Ashwinkumar:2024vys, Ishtiaque:2024orn}. It would be interesting to understand how these results generalize in the presence of matter fields, and for supergroup gauge groups.

Finally, it would be interesting to understand the relationship between the 4d chiral WZW model and the well-known 4d WZW model studied, for example, in \cite{Donaldson:1985zz, Nair:1990aa,Losev:1995cr}.
The latter has recently been studied in the context of  celestial holography, where the semiclassical bulk spacetime is a 4d asymptotically flat, self-dual Kähler geometry, known  as Burns space \cite{Costello:2023hmi}. Given that 5d non-commutative Chern-Simons theory is related to a 5d theory that generalizes 5d K\"ahler Chern-Simons theory via a Seiberg-Witten transform \cite{Aganagic}, we expect the 4d chiral WZW model we introduced in this work and the well-known (K\"ahler) 4d WZW model to be related via a Seiberg-Witten transform. 
 \newline
\newline
\noindent \textbf{Acknowledgements}
We would like to thank Matthias Blau, Kevin Costello, Nafiz Ishtiaque, Jihwan Oh, Meng-Chwan Tan, Konstantin Wernli and Masahito Yamazaki for helpful comments and discussions. Kavli IPMU is supported by the World Premier International Research Center Initiative (WPI), MEXT, Japan.
DJS was supported in part by the STFC Consolidated grant ST/T000708/1.

\appendix
\section{Gauge Invariance of Matter Action }\label{matgaug}
With non-commutativity turned off, the 5d matter action is 
\ie
&\int_{\mathbb{R} \times \mathbb{C}^2} dz\wedge dw \wedge \eta \wedge d_A\phi\\=& \int_{\mathbb{R} \times \mathbb{C}^2} dz\wedge dw \wedge dt \wedge  d\bar{z} \wedge d\bar{w} \textrm{ } \varepsilon^{ijk }\eta_{iI} (D_j \phi_k)^I
\\=& \int_{\mathbb{R} \times \mathbb{C}^2} dz\wedge dw \wedge dt \wedge  d\bar{z} \wedge d\bar{w} \textrm{ } \varepsilon^{ijk }\eta_{iI} (\partial_j \phi^I_k + A_j^a \rho(t_a)^I{ }_J\phi^J_k ),
\fe
where the index $I$ runs over $1,\ldots, D_R$, where $D_R$ is the dimension of the representation $R$ and its conjugate.
Writing $dz\wedge dw \wedge dt \wedge  d\bar{z} \wedge d\bar{w}$ as $d^5x$, the action transforms under the gauge transformations 
\ie 
\delta \eta_{iI}= -\epsilon^a \eta_{iJ} \rho\left(t_a\right)^J{ }_I, \quad \delta \phi_k^I=\epsilon^a \rho\left(t_a\right)^I{ }_J \phi_k^J, \quad \delta A^a=-D \epsilon^a = - d\epsilon^a - f^{abc}A_b \epsilon_c,
\fe
as
\ie 
\begin{aligned} & \int d^5x \varepsilon^{ijk} \left(-\epsilon^a \eta_{iI} \rho\left(t_a\right)^I{ }_J \partial_j \phi_k^J+\eta_{iI} \partial_j\left(\epsilon^a \rho\left(t_a\right)^I{ }_J \phi_k^J\right)-\partial_j \epsilon^a \eta_{iI} \rho\left(t_a\right)^I{ }_J \phi_k^J\right. \\ & \left.-A_j^a \epsilon^b \eta_{iI} \rho\left(t_b\right)^I{ }_J \rho\left(t_a\right)^J{ }_K \phi_k^K+A_j^a \epsilon^b \eta_{iI} \rho\left(t_a\right)^I{ }_J\rho\left(t_b\right)^J{ }_K \phi_k^K-f^{cab  } A_j^a \epsilon_b \eta_{iI} \rho\left(t_c\right)^I{ }_J \phi_k^J\right)=0,\end{aligned}
\fe 
where we have used the Lie algebra relation $\rho(t_a)\rho(t_b)-\rho(t_b)\rho(t_a)=f_{ab}^c\rho(t_c)$.

In the non-commutative case, the matter action is 
\ie
\int_{\mathbb{R} \times \mathbb{C}^2} dz\wedge dw \wedge dt \wedge  d\bar{z} \wedge d\bar{w} \textrm{ } \varepsilon^{ijk }\eta_{iI} (\partial_j \phi^I_k + (A_j)^I{ }_J*\phi^J_k ), 
\fe 
and the gauge transformations are 
\ie 
\delta \eta_{iI}= -\eta_{iJ}*\epsilon^J{ }_I   \, \quad \delta \phi_k^I=\epsilon^I{ }_J *\phi_k^J, \quad \delta A=-D \epsilon = - d\epsilon - A *\epsilon + \epsilon *A.
\fe
The gauge transformation of the non-commutative matter action is
\ie 
\begin{aligned} & \int d^5x \varepsilon^{ijk} \left(-\eta_{iJ}*\epsilon^J{ }_I \partial_j \phi_k^I+\eta_{iI} \partial_j\left(\epsilon^I{ }_J *\phi_k^J\right) -\eta_{iJ}*\epsilon^J{ }_I (A_j)^I{ }_K  *\phi_k^K \right. \\ & \left. 
+\eta_{iI}(-\partial_j \epsilon^I{ }_J- (A_j*\epsilon)^I{ }_J + (\epsilon * A_j)^I{ }_J  )\phi_k^J
 +\eta_{iI} (A_j)^I{ }_J*(\epsilon^J{ }_K*\phi_k^K) \right)\\&=0,\end{aligned}
\fe 
which follows from the Moyal product identities
\ie
\int d^5x  (A*BC)=
\int d^5x (AB*C)
\fe
and 
\ie
\int d^5x  (A*BC*D)=
\int d^5x (AB*C*D).
\fe

\section{Computation of Finite Gauge Variation of Non-commutative 5d CS Action}
To compute the variation of the 5d non-commutative CS action 
\begin{eqnarray} 
\int_{\mathbb{R}_+ \times \mathbb{C}^2 } dz \wedge dw \wedge  CS(A).
\end{eqnarray} 
We consider the gauge variation of the Chern-Simons functional
under finite gauge transformations of the form 
\begin{equation} \label{gt}
A=-dg*g^{-1}+g *\tilde{A}* g^{-1}.
\end{equation} 
To compute this, we define
$$
\widehat{A}:=-d {g} *{g}^{-1}, \quad A^{\prime}:={g}* \tilde{A} *{g}^{-1}.
$$
For the sum of two connections, the Chern-Simons functional can be written as
\begin{eqnarray}
\begin{aligned}
C S(A) & =\int \textrm{Tr} \bigg((\widehat{A}+A')\wedge_* d(\widehat{A}+A')+\frac{2}{3}(\widehat{A}+A')\wedge_*  (\widehat{A}+A')\wedge_* (\widehat{A}+A') \bigg)\\
& =\int \textrm{Tr} \bigg(\widehat{A}\wedge * d\widehat {A}+\widehat{A}\wedge_* dA' +A'\wedge_*  d\widehat{A} +A' \wedge_* dA' + \frac{2}{3}\widehat{A}\wedge * \widehat{A}\wedge \widehat{A} \\
&  +2 A'\wedge_* \widehat{A} \wedge_* \widehat{A} + 2 A'\wedge_* A' \wedge_* \widehat{A} +  \frac{2}{3}A'\wedge_* A'\wedge_* A'
\bigg)\\
&={CS}(\widehat{A})+2 \operatorname{Tr}\left(F(\widehat{A}) \wedge_* A^{\prime}\right)-\mathrm{d} \operatorname{Tr}\left(\widehat{A} \wedge_* A^{\prime}\right)+2 \operatorname{Tr}\left(\widehat{A} \wedge_* A^{\prime} \wedge_* A^{\prime}\right)+{CS}\left(A^{\prime}\right),
\end{aligned} \label{csga0}
\end{eqnarray}.
The third term
 is a total derivative that takes the form 
 \begin{equation} 
- d \textrm{Tr}\left(\widehat{A}\wedge * A^{\prime}\right) = \textrm{Tr }d (g^{-1} * dg \wedge_* \tilde{A})
 \end{equation}
 In addition, we find that 
\begin{eqnarray} 
\begin{aligned}
C S\left(A^{\prime}\right) & = \textrm{Tr} \left( g*\tilde{A} *g^{-1} \wedge_* d (g *\tilde{A} *g^{-1}) + \frac{2}{3} g * \tilde{A}* g^{-1} \wedge_* g *\tilde{A} *g^{-1} \wedge_* g *\tilde{A} g^{-1}  \right) \\
&= \textrm{Tr} \left( CS(\tilde{A} ) +2 \tilde{A} \wedge_* \tilde{A} \wedge_* g^{-1}* d g\right),
\end{aligned}
\end{eqnarray} 
the second term of which cancels
\begin{eqnarray} 
2 \operatorname{Tr}\left(\widehat{A} \wedge_* A^{\prime} \wedge_* A^{\prime}\right) = \textrm{Tr} (2 g *\tilde{A} *g^{-1} \wedge_* g * \tilde{A} * g^{-1} \wedge_* (-dg*g^{-1} )) \nonumber \\ = - \textrm{Tr} (2 \tilde{A} \wedge_* \tilde{A} \wedge_* g^{-1} * d g ).
\end{eqnarray}
Finally, the Chern-Simons functional for a gauge field that is pure gauge takes the form 
\begin{equation}
C S(\widehat{A})= -\frac{1}{3}\left(\widehat{A} \wedge_* \widehat{A} \wedge_*\widehat{A}\right) = \frac{1}{3}\textrm{Tr} (g^{-1} * dg \wedge_*g^{-1} * dg \wedge_* g^{-1} * dg  ),
\end{equation}
using $F(\widehat{A})=d\widehat{A} + \widehat{A} \wedge_* \widehat{A} =0 $.

Thus, we find that 
\begin{eqnarray} 
CS(A)=CS(\tilde{A})+ \textrm{Tr }d (g^{-1} * dg \wedge_* \tilde{A}) + \frac{1}{3}\textrm{Tr} (g^{-1} * dg \wedge_* g^{-1} * dg \wedge_* g^{-1} *dg  ).
\end{eqnarray}

\section{Details of BV-BFV Formalism Computation}\label{B}

In what follows, we shall evaluate the Feynman diagrams that contribute to the effective action of 5d CS on $I\times \C^2$, as depicted in Figure \ref{feyn}. For concision, we shall suppress factors of $\hbar$  :
\newline
\newline 
(i)
$\int_{\C^2}dz\wedge dw\wedge  \left\langle A_{\text {out }}^1,\left.a^2\right|_{t=1}\right\rangle \int_{\C^2}dz'\wedge dw'\wedge  \left\langle\left. a^1\right|_{t=0}, A_{\text {in }}^2\right\rangle=\int_{\C^2}dz\wedge dw\wedge   \left\langle\mathrm{A}_{\text {out }}^1, e^{-\mathrm{ad}_\sigma}  \mathrm{A}_{\text {in }}^2\right\rangle$. 
\newline
\newline
(ii)
\ie
\begin{aligned}
& -\int_{\C^2}dz\wedge dw \wedge\left\langle\mathrm{A}_{\text {out }}^{1},\left.a^{2}\right|_{t=1}\right\rangle \int_{I \times \C^2 } dz'\wedge dw'\wedge d t'\left\langle a^{1}, \bar{\partial}^2 \sigma\right\rangle\\&=-\int_{\C^2}dz\wedge dw\wedge \left\langle\mathrm{A}_{\text {out }}^{1}, \int_0^1 d t e^{-(1-t) \mathrm{ad}_\sigma} \bar{\partial}^2 \sigma\right\rangle \\&=-\int_{\C^2}dz\wedge dw\wedge  \left\langle\mathrm{A}_{\text {out }}^1, \frac{1-e^{-\mathrm{ad}_\sigma}}{\operatorname{ad}_\sigma}  \bar{\partial} ^2\sigma\right\rangle . 
&
\end{aligned}
\fe
\newline
\newline
(iii)
\ie 
\begin{aligned}
 \int_{I \times \C^2} dz\wedge dw \wedge d t\left\langle\bar{\partial}^1 \sigma, a^2\right\rangle \int_{\C^2} dz'\wedge dw'\left\langle\left. a^1\right|_{t=0}, \mathrm{~A}_{\mathrm{in}}^2\right\rangle= \int_{\C^2}dz\wedge dw \wedge \left\langle \bar{\partial}^1 \sigma, \int_0^1 d t e^{-t\operatorname{ad}_\sigma} \mathrm{A}_{\mathrm{in}}^2\right\rangle
\nonumber
\end{aligned}
\fe 
$$
\begin{aligned}
= & \int_{\C^2}dz\wedge dw \wedge\left\langle\bar{\partial}^1 \sigma, \frac{1-e^{-\mathrm{ad}_\sigma}}{\mathrm{ad}_\sigma}  \mathrm{A}_{\mathrm{in}}^2\right\rangle=\int_{\C^2}dz\wedge dw \wedge\left\langle\frac{e^{\mathrm{ad}_\sigma}-1}{\mathrm{ad}_\sigma} \bar{\partial}^1 \sigma, \mathrm{A}_{\mathrm{in}}^2\right\rangle \\
\text { (iv) } & -\int_{I \times \C^2} dz\wedge dw \wedge d t \langle \bar{\partial}^1 \sigma, a^2 \rangle \int_{I \times \C^2}dz'\wedge dw' \wedge d t^{\prime}\langle a^1, \bar{\partial}^2 \sigma \rangle\\=&-\int_{\C^2} \int_{\C^2}dz\wedge dw \wedge\langle \bar{\partial}^1 \sigma, \int_0^1 d t \int_0^t d t^{\prime} e^{- (t-t^{\prime} ) \operatorname{ad}_\sigma}  \bar{\partial}^2 \sigma \rangle \\
= & -\int_{\C^2} dz\wedge dw \wedge\langle\bar{\partial}^1 \sigma, \frac{e^{-\mathrm{ad}_\sigma}+\operatorname{ad}_\sigma-1}{(\operatorname{ad}_\sigma )^2} \bar{\partial}^2 \sigma \rangle \\
\text { (v) } & -\int_{I \times \C^2} dz\wedge dw \wedge d t\left\langle\operatorname{ad}_\sigma \mathrm{A}_{\mathrm{res}}^*, c_{\mathrm{fl}}\right\rangle \int_{\C^2}dz'\wedge dw' \wedge\left\langle\left.\mathrm{A}_{\mathrm{fl}}^*\right|_{t=1}, c_{\mathrm{out}}\right\rangle + \int_{\C^2}dz\wedge dw \wedge\left\langle \mathrm{A}_{\mathrm{res}}^*, c_{\mathrm{out}}\right\rangle \\
= & -\int_{\C^2}dz\wedge dw \wedge\left\langle\operatorname{ad}_\sigma \mathrm{A}_{\mathrm{res}}^*, \int_0^1 d t \frac{e^{(1-t) \mathrm{ad}_\sigma}-e^{\mathrm{ad}_\sigma}}{e^{\mathrm{ad}_\sigma}-1} c_{\mathrm{out}}\right\rangle+\int_{\C^2}dz\wedge dw \wedge\left\langle \mathrm{A}_{\mathrm{res}}^*, c_{\mathrm{out}}\right\rangle \\
= & -\int_{\C^2}dz\wedge dw \wedge\left\langle\mathrm{A}_{\mathrm{res}}^*, \frac{\mathrm{ad}_\sigma}{1-e^{-\mathrm{ad}_\sigma} } c_{\mathrm{out}} \right\rangle \\
\text { (vi) }
&  \int_{I \times \C^2} dz\wedge dw \wedge d t\left\langle\operatorname{ad}_\sigma \mathrm{A}_{\mathrm{res}}^*, c_{\mathrm{fl}}\right\rangle \int_{\C^2}dz'\wedge dw' \wedge\left\langle\left.\mathrm{A}_{\mathrm{fl}}^*\right|_{t=0}, c_{\mathrm{in}}\right\rangle - \int_{\C^2}dz\wedge dw \wedge\left\langle \mathrm{A}_{\mathrm{res}}^*, c_{\mathrm{in}}\right\rangle \\
& =\int_{\C^2}dz\wedge dw \wedge\left\langle\operatorname{ad}_\sigma \mathrm{A}_{\mathrm{res}}^*, \int_0^1 d t \frac{e^{(1-t) \operatorname{ad}_\sigma}-1}{e^{\mathrm{ad}_\sigma}-1} c_{\mathrm{in}}\right\rangle -\int_{\C^2}dz\wedge dw \wedge\left\langle \mathrm{A}_{\mathrm{res}}^*, c_{\mathrm{in}}\right\rangle \\
& =\int_{\C^2}dz\wedge dw \wedge\left\langle\mathrm{A}_{\mathrm{res}}^*, \frac{\mathrm{ad}_\sigma}{e^{\mathrm{ad}_\sigma}-1}c_{\mathrm{in}}\right\rangle
\end{aligned}
$$
One can further show that (vii) is zero since
\ie
\int_{t_1 \ldots \ldots . t_k \in[0.1]} d t_1 \cdots d t_k \theta\left(t_1-t_2\right) \theta\left(t_2-t_3\right) \cdots \theta\left(t_{k-1}-t_k\right) \theta\left(t_k-t_1\right)=0
\fe
Finally (viii) can be evaluated to be 
\ie
-i \hbar \mathbb{W}=-i \hbar \sum_{p \in (\C\times \C)^{(reg)}} j(\sigma(p))
\fe
where $(\C\times \C)^{\textit{(reg)}}$ is a regularization of $\C^2$ consisting of a finite set of points and 
\ie 
j(\sigma)=\operatorname{tr}_{{G}} \log \frac{\sinh \frac{\mathrm{ad}_\sigma}{2}}{\frac{\mathrm{ad}_\sigma}{2}}.
\fe 
The quantity $i \hbar \mathbb{W}$ is a divergent term that can be interpreted as a change of path integral measure, see \cite{Cattaneo:2020lle}.

Defining $F_+$ and $F_-$ as in \eqref{fpm}, we thus find that the effective action can be rewritten as 
\ie\begin{aligned}  &S^{\mathrm{eff}} \\  \quad=&\int_{\C^2}dz\wedge dw \bigg(\left\langle\mathrm{A}_{\text {out }}^1, e^{-\mathrm{ad}_\sigma}  \mathrm{A}_{\mathrm{in}}^2\right\rangle - \left\langle\mathrm{A}_{\mathrm{out}}^1, \frac{1-e^{-\mathrm{ad}_\sigma}}{\mathrm{ad}_\sigma}  \bar{\partial}^2 \sigma\right\rangle -\left\langle\mathrm{A}_{\mathrm{in}}^2, \frac{e^{\mathrm{ad}_\sigma}-1}{\mathrm{ad}_\sigma} \bar{\partial}^1 \sigma\right\rangle \\ & -\left\langle\bar{\partial}^1 \sigma, \frac{e^{-\mathrm{ad}_\sigma}+\mathrm{ad}_\sigma-1}{\left(\operatorname{ad}_\sigma\right)^2}  \bar{\partial}^2 \sigma\right\rangle-\left\langle\mathrm{A}_{\mathrm{res}}^*, F_{+}\left(\operatorname{ad}_\sigma\right)  c_{\mathrm{out}}+F_{-}\left(\operatorname{ad}_\sigma\right) c_{\mathrm{in}}\right\rangle\bigg)-i \hbar \mathbb{W}(\sigma) .\end{aligned}\fe

We shall now explain how the effective action can be further reexpressed as 
$$\begin{aligned}  S^{\mathrm{eff}}=&\int_{\C^2}dz \wedge dw\left(\left\langle\mathrm{A}_{\text {out }}^1, g \mathrm{~A}_{\mathrm{in}}^2 g^{-1}\right\rangle + \left\langle\mathrm{A}_{\text {out }}^1, \bar{\partial}^2 g \cdot g^{-1}\right\rangle + \left\langle\mathrm{A}_{\text {in }}^2, g^{-1} \bar{\partial}^1 g\right\rangle\right. \\ & \left.\quad-\left\langle\mathrm{A}_{\text {res }}^*, F_{+}\left(\operatorname{ad}_{-\log g}\right)  c_{\text {out }}+F_{-}\left(\operatorname{ad}_{-\log g}\right) c_{\text {in }}\right\rangle\right)+\mathrm{WZW}(g)-i \hbar \mathbb{W} .   
\end{aligned}$$
$\begin{aligned}
\mathrm{WZW}(g)=&-\frac{1}{2} \int_{\C^2}dz \wedge dw \wedge \left\langle\bar{\partial}^1 g \cdot g^{-1}, \bar{\partial}^2 g \cdot g^{-1}\right\rangle\\& - \frac{1}{12} \int_{I \times \C^2}dz \wedge dw \wedge \left\langle d \widetilde{g} \cdot \widetilde{g}^{-1},\left[d \widetilde{g} \cdot \widetilde{g}^{-1}, d \widetilde{g} \cdot \widetilde{g}^{-1}\right]\right\rangle .\end{aligned}$

Firstly, note that
$$\begin{aligned} g^{-1} \partial_{\bar{z}} g=e^\sigma \int_0^1 d \tau e^{-\tau \sigma}(-\partial_{\bar{z}} \sigma) e^{-(1-\tau) \sigma} & =\int_0^1 d \tau e^{\tau \operatorname{ad}_\sigma}(-\partial_{\bar{z}} \sigma) \\ & =\frac{e^{\operatorname{ad}_\sigma}-1}{\operatorname{ad}_\sigma}(-\partial_{\bar{z}} \sigma), \\ {\partial_{\bar{w}}} g \cdot g^{-1}=\int_0^1 d \tau e^{-\tau \sigma}(-{\partial_{\bar{w}}} \sigma) e^{-(1-\tau) \sigma} e^\sigma & =\int_0^1 d \tau e^{-\tau \operatorname{ad}_\sigma}(-{\partial_{\bar{w}}} \sigma) \\ & =\frac{1-e^{-\operatorname{ad}_\sigma}}{\operatorname{ad}_\sigma}(-{\partial_{\bar{w}}} \sigma) .\end{aligned}$$

The Wess-Zumino term can be rewritten as 
\ie \begin{aligned} & \quad-\frac{1}{12} \int_{I \times \C^2}dz\wedge dw \left\langle d \widetilde{g} \cdot \widetilde{g}^{-1},\left[d \widetilde{g} \cdot \widetilde{g}^{-1}, d \widetilde{g} \cdot \widetilde{g}^{-1}\right]\right\rangle \\ & =-\frac{1}{4} \int_{{\C^2}} dz \wedge dw \wedge \\ &\; \;  \; \;  \int_0^1 d t\left\langle\sigma, \int_0^{1-t} d \tau \int_0^{1-t} d \tau^{\prime}\left[e^{-\tau \sigma}(-d \sigma) e^{-(1-t-\tau) \sigma} e^{(1-t) \sigma}, e^{-\tau^{\prime} \sigma}(-d \sigma) e^{-\left(1-t-\tau^{\prime}\right) \sigma} e^{(1-t) \sigma}\right]\right\rangle \\ & =\frac{1}{4} \int_{\C^2} dz \wedge dw \wedge  \int_0^1 d t \int_0^{1-t} d \tau \int_0^{1-t} d \tau^{\prime}\left\langle d \sigma,\left[\sigma, e^{\left(\tau^{\prime}-\tau\right) \operatorname{ad}_\sigma} d \sigma\right]\right\rangle\\
&=\frac{1}{2} \int_{\C^2} dz \wedge dw \wedge \left\langle d \sigma,\left[\sigma,\left(\frac{\sinh ^2 \operatorname{ad}_\sigma-\operatorname{ad}_\sigma}{\left(\operatorname{ad}_\sigma\right)^3}\right) d \sigma\right]\right\rangle\\
&=\frac{1}{2} \int_{\C^2}dz\wedge dw \wedge \left\langle d \sigma,\left(\frac{\sinh \operatorname{ad}_\sigma-\operatorname{ad}_\sigma}{\left(\operatorname{ad}_\sigma\right)^2}\right) d \sigma\right \rangle 
\\&=  \int_{\C^2}dz\wedge dw \wedge \left\langle \bar{\partial}^1 \sigma,\left(\frac{\sinh \operatorname{ad}_\sigma-\operatorname{ad}_\sigma}{\left(\operatorname{ad}_\sigma\right)^2}\right) \bar{\partial}^2 \sigma\right
\rangle 
\end{aligned}\fe 

The WZW kinetic term is:
$$
 \begin{aligned}
& -\frac{1}{2} \int_{\C^2} dz \wedge dw \wedge \left\langle\bar{\partial}^1 g \cdot g^{-1}, \bar{\partial}^2 g \cdot g^{-1}\right\rangle \\
= & -\frac{1}{2} \int_{\C^2}dz \wedge dw \wedge  \int_0^1 d \tau \int_0^1 d \tau^{\prime}\left\langle e^{-\tau \sigma}(-\bar{\partial}^1 \sigma) e^{-(1-\tau) \sigma} e^\sigma, e^{-\tau^{\prime} \sigma}(-\bar{\partial}^2 \sigma) e^{-\left(1-\tau^{\prime}\right) \sigma} e^\sigma\right\rangle \\
 =&-\frac{1}{2} \int_{\C^2}dz \wedge dw \wedge  \int_0^1 d \tau \int_0^1 d \tau^{\prime}\left\langle\bar{\partial}^1 \sigma, e^{\left(\tau-\tau^{\prime}\right) \operatorname{ad}_\sigma} \bar{\partial}^2 \sigma\right\rangle \\=&-\int_{\C^2}dz \wedge dw \wedge  \left\langle\bar{\partial}^1 \sigma, \frac{\cosh _\sigma-1}{\left(\operatorname{ad}_\sigma\right)^2}\bar{\partial}^2 \sigma\right\rangle.
\end{aligned}
$$
We thus find that the 4d WZW action is 
$$\begin{aligned} \mathrm{WZW}(g)=\int_{\C^2} dz \wedge dw \left(-\left\langle\bar{\partial}^1 \sigma, \frac{\cosh \operatorname{ad}_\sigma-1}{\left(\operatorname{ad}_\sigma\right)^2} \bar{\partial}^2 \sigma\right\rangle+\left\langle\bar{\partial}^1 \sigma, \frac{\sinh \operatorname{ad}_\sigma-\operatorname{ad}_\sigma}{\left(\operatorname{ad}_\sigma\right)^2} \bar{\partial}^2 \sigma\right\rangle \right) \\ \\ =-\int_{\C^2} dz \wedge dw \wedge \left\langle\bar{\partial}^1 \sigma, \frac{e^{-\operatorname{ad}_\sigma}+\operatorname{ad}_\sigma-1}{\left(\operatorname{ad}_\sigma\right)^2} \bar{\partial}^2 \sigma\right\rangle .\end{aligned}$$

In this way, we have shown that the 5d CS path integral is equal to the path integral of a gauged 4d chiral WZW model with an action of the form 
\ie 
\begin{aligned}
S^{\mathrm{eff}}=&\int_{\C^2}dz \wedge dw\left(\left\langle\mathrm{A}_{\text {out }}^1, g \mathrm{~A}_{\mathrm{in}}^2 g^{-1}\right\rangle + \left\langle\mathrm{A}_{\text {out }}^1, \bar{\partial}^2 g \cdot g^{-1}\right\rangle + \left\langle\mathrm{A}_{\text {in }}^2, g^{-1} \bar{\partial}^1 g\right\rangle\right. \\ & \left.\quad-\left\langle\mathrm{A}_{\text {res }}^*, F_{+}\left(\operatorname{ad}_{-\log g}\right)  c_{\text {out }}+F_{-}\left(\operatorname{ad}_{-\log g}\right) c_{\text {in }}\right\rangle\right)+\mathrm{WZW}(g)-i \hbar \mathbb{W}.
\end{aligned}
\fe 
 
\subsection{Non-commutative generalization}\label{ncgen}

Let us now explain how to generalize the computation above to the non-commutative case, first focusing on $U(1)$ 5d Chern-Simons theory. To this end, we need to show that the exponential map defined using the Moyal product, 
\ie 
g=e^{\phi}_*=1+ \phi + \frac{1}{2} \phi * \phi + \frac{1}{3!} \phi * \phi * \phi + \ldots 
\fe 
obeys a version of Duhamel's formula for its derivative :
\ie \label{dernc}
e_*^{-X}  *\frac{d}{dt} e^{X(t)}_* = \int_0^1 e^{- s\textrm{ ad}_* X } ds * \frac{dX}{dt}. 
\fe 
To achieve this, we first note that 
\ie 
e^X_* &= \lim_{N\rightarrow \infty } \left(1+ \frac{X}{N}\right)* \left(1+ \frac{X}{N}\right) * \ldots  \textrm{ ($N$ times)}\\
& = \lim_{N\rightarrow \infty } \left(1+ \frac{X}{N}\right)_*^N
\fe
Taking the derivative with respect to a parameter $t$ on which $X$ depends, we have 
\ie 
\frac{d}{dt} e^X_*  = &\lim_{N \rightarrow \infty }  \sum^N_{k=1} \left( 1+ \frac{X(t)}{N} \right)_*^{N-k} * \frac{1}{N} \frac{dX}{dt }* \left( 1+ \frac{X(t)}{N} \right)_*^{k-1} \\
=&\int_0^1 e_*^{(1-s)X} * \frac{dX}{dt} * e_*^{sX }ds
\fe
This is equivalent to \eqref{dernc} if 
\ie 
\textrm{Ad}_{*} e_*^{-sX} = e^{- s\textrm{ ad}_* X }, 
\fe 
which holds since 
\ie 
\frac{d}{ds} \left( e_*^{sX} * Y *  e_*^{-sX}\right) &= X* e_*^{sX} * Y *  e_*^{-sX} - e_*^{sX} * Y *  e_*^{-sX} *X \\
&= \textrm{ad}_*X \left( e_*^{sX} * Y *  e_*^{-sX}\right).
\fe 

Thus, the computations in the previous subsection go through if we replace the Lie algebra commutators by Moyal bracket commutators. These results naturally extend to the non-commutative version of the gauge group $U(N)$ as well.

\subsection{4d WZW-Matter Theory from Feynman Diagrams}\label{4dwzm} 
In this section, we explore the detailed computation of tree-level Feynman diagrams for the 4D Wess-Zumino-Witten (WZW)-matter theory, as derived from  5d Chern-Simons-matter theory in the BV-BFV formalism. For concision, we shall suppress factors of $\hbar$. 

The relevant Feynman diagrams are depicted in Figure \ref{feyn}.
Feynman diagram (i) is evaluated as 
    \begin{equation}
  \int_{\mathbb{C}^2} dz \wedge dw \wedge  \langle  \left.\tilde{\phi}_{\text{out}}^1 , a_{{\eta}}^2 \right|_{t=1} \rangle \int_{\mathbb{C}^2} dz' \wedge dw' \wedge  \langle \left. a_{{\phi}}^1 \right|_{t=0}, \tilde{\eta}_{\text{in}}^2 \rangle 
      = \int_{\mathbb{C}^2} dz \wedge dw \wedge \langle \tilde{\phi}_{\text{out}}^1, e^{-\operatorname{ad}_{-\log g}}  \tilde{\eta}_{\text{in}}^2 \rangle,
    \end{equation}
     Feynman diagram (ii) is evaluated as 
    \begin{equation}
    \begin{aligned}
    & \int_{\{1\}\times \mathbb{C}^2} dz \wedge dw \wedge \langle \tilde{\phi}_{\text{out}}^1, \left. a_{{\eta}}^2 \right|_{t=1} \rangle \int_{I \times \mathbb{C}^2} dz' \wedge dw' \wedge dt' \wedge \, \langle a_{{\phi}}^1, {\partial}^2 \sigma_{{\eta}} \rangle \\
    & = \int_{\mathbb{C}^2} dz \wedge dw \wedge\left\langle {\tilde{\phi}}_{\text{out}}^1, \int_0^1 dt \, e^{-(1-t) \operatorname{ad}_{-\log g}} {\partial}^2 \sigma_{{\eta}} \right\rangle \\
    & = \int_{\mathbb{C}^2} dz \wedge dw \wedge \left\langle \tilde{\phi}_{\text{out}}^1, \frac{1 - e^{-\operatorname{ad}_{-\log g}}}{\operatorname{ad}_{-\log g}}{\partial}^2 \sigma_{{\eta}} \right\rangle,
    \end{aligned}
\end{equation}
      Feynman diagram (iii) is evaluated as 
   \ie
    \begin{aligned}
    &- \int_{I \times \mathbb{C}^2} dz \wedge dw \wedge dt\wedge \, \langle {\partial}^1 \sigma_{{\phi}}, a_{{\eta}}^2 \rangle \int_{\mathbb{C}^2} dz'\wedge dw'\langle \left. a_{{\phi}}^1 \right|_{t=0}, {\eta}_{\text{in}}^2 \rangle \\
    &=  -\int_{\mathbb{C}^2} dz \wedge dw \wedge  \left\langle {\partial}^1 \sigma_{{\phi}}, \int_0^1 dt \, e^{-t \operatorname{ad}_{-\log g}} {\eta}_{\text{in}}^2 \right\rangle \\
    &=- \int_{\mathbb{C}^2} dz \wedge dw \wedge  \left\langle {\partial}^1 \sigma_{{\phi}}, \frac{1 - e^{-\operatorname{ad}_{-\log g}}}{\operatorname{ad}_{-\log g}} {\eta}_{\text{in}}^2 \right\rangle ,
    \end{aligned}
\fe
     Feynman diagram (iv) is evaluated as 
 \ie
    \begin{aligned}
   & - \int_{I \times \mathbb{C}^2} dz \wedge dw \wedge dt \wedge \, \langle {\partial}^1 \sigma_{{\eta}}, a_{{\phi}}^2 \rangle \int_{I \times \mathbb{C}^2} dz' \wedge dw' \wedge dt' \wedge \mathrm{Tr} \, \langle a_{{\eta}}^1, {\partial}^2 \sigma_{{\phi}} \rangle \\
    &= - \int_{\mathbb{C}^2} dz \wedge dw \wedge\left\langle {\partial}^1 \sigma_{{\phi}}, \int_0^1 dt \int_0^t dt' \, e^{-(t - t') \operatorname{ad}_{-\log g}} {\partial}^2 \sigma_{{\eta}} \right \rangle \\
    &= - \int_{\mathbb{C}^2} dz \wedge dw \wedge\left \langle {\partial}^1 \sigma_{{\phi}}, \frac{e^{-\operatorname{ad}_{-\log g}} + \operatorname{ad}_{-\log g} - 1}{(\operatorname{ad}_{-\log g})^2} {\partial}^2 \sigma_{{\eta}} \right \rangle,
    \end{aligned}
    \fe
  Feynman diagram (v) is evaluated as 
     \ie
    \begin{aligned}
   &  -  \int_{I \times \mathbb{C}^2} dz \wedge dw \wedge dt \wedge \, \langle {\partial}^1 \sigma_{{\phi}}, a_{{\eta}}^2 \rangle \int_{I \times \mathbb{C}^2} dz' \wedge dw' \wedge dt' \wedge \, \langle a_{{\phi}}^1, {\partial}^2 \sigma_{{\eta}} \rangle \\
    &= -  \int_{\mathbb{C}^2} dz \wedge dw \wedge \left\langle {\partial}^1 \sigma_{{\eta}}, \int_0^1 dt \int_0^t dt' \, e^{-(t - t') \operatorname{ad}_{-\log g}} {\partial}^2 \sigma_{{\phi}} \right \rangle \\
    &=  -\int_{\mathbb{C}^2} dz \wedge dw \wedge \left \langle {\partial}^1 \sigma_{{\eta}}, \frac{e^{-\operatorname{ad}_{-\log g}} + \operatorname{ad}_{-\log g} - 1}{(\operatorname{ad}_{-\log g})^2} {\partial}^2 \sigma_{{\phi}} \right \rangle,
    \end{aligned}
    \fe
  Feynman diagram (vi) is evaluated as 
   \ie
    \begin{aligned}\label{pq6}
    & \int_{I \times \mathbb{C}^2} dz \wedge dw \wedge dt\wedge  \, \langle \operatorname{ad}_{\sigma_{\phi}} {\phi}_{\text{res}}^*-\operatorname{ad}_{\sigma_{\eta}} {\eta}_{\text{res}}^*, c_{\text{fl}} \rangle \int_{\mathbb{C}^2} dz' \wedge dw' \wedge \langle \left. \mathrm{A}_{\text{fl}}^* \right|_{t=0}, c_{\text{in}} \rangle 
    \\
    &=  \int_{\mathbb{C}^2} dz \wedge dw \wedge  \left\langle \operatorname{ad}_{\sigma_{\phi}} {\phi}_{\text{res}}^*-\operatorname{ad}_{\sigma_{\eta}} {\eta}_{\text{res}}^*, \int_0^1 dt \, \frac{e^{(1 - t) \operatorname{ad}_{-\log g}} - e^{\operatorname{ad}_{-\log g}}}{e^{\operatorname{ad}_{-\log g}} - 1} c_{\text{in}} \right\rangle \\
    &=  \int_{\mathbb{C}^2} dz \wedge dw \wedge \left\langle \operatorname{ad}_{\sigma_{\phi}} {\phi}_{\text{res}}^*-\operatorname{ad}_{\sigma_{\eta}} {\eta}_{\text{res}}^*, \left( \frac{1}{\textrm{ad}_{-\log g}} -\frac{1}{1-e^{-\textrm{ad}_{-\log g}}}\right) c_{\text{in}} \right \rangle,
    \end{aligned}
\fe
    and Feynman diagram (vii) is evaluated as 
    \ie
    \begin{aligned}\label{pq5}
    & -\int_{I \times \mathbb{C}^2} dz \wedge dw \wedge dt\wedge  \, \langle \operatorname{ad}_{\sigma_{\phi}} {\phi}_{\text{res}}^*-\operatorname{ad}_{\sigma_{\eta}} {\eta}_{\text{res}}^*, c_{\text{fl}} \rangle \int_{\mathbb{C}^2} dz' \wedge dw' \wedge \langle \left. \mathrm{A}_{\text{fl}}^* \right|_{t=1}, c_{\text{out}} \rangle 
    \\
    &=  -\int_{\mathbb{C}^2} dz \wedge dw \wedge  \left\langle \operatorname{ad}_{\sigma_{\phi}} {\phi}_{\text{res}}^*-\operatorname{ad}_{\sigma_{\eta}} {\eta}_{\text{res}}^*, \int_0^1 dt \, \frac{e^{(1 - t) \operatorname{ad}_{-\log g}} - e^{\operatorname{ad}_{-\log g}}}{e^{\operatorname{ad}_{-\log g}} - 1} c_{\text{out}} \right\rangle \\
    &=  -\int_{\mathbb{C}^2} dz \wedge dw \wedge \left\langle \operatorname{ad}_{\sigma_{\phi}} {\phi}_{\text{res}}^*-\operatorname{ad}_{\sigma_{\eta}} {\eta}_{\text{res}}^*, \left( \frac{1}{\textrm{ad}_{-\log g}} -\frac{1}{1-e^{-\textrm{ad}_{-\log g}}}\right) c_{\text{out}} \right \rangle.
    \end{aligned}
\fe

There are also several wheel diagrams of the form given in Feynman diagram (viii), which can be shown to evaluate to zero, as in the computation for 5d Chern-Simons theory without matter.

\newpage

\bibliographystyle{ytphys.bst}
\bibliography{5dCSmatter}

\providecommand{\href}[2]{#2}\begingroup\raggedright\begin{thebibliography}{10}

\bibitem{Costello:2016nkh}
K.~Costello, ``{M-theory in the Omega-background and 5-dimensional non-commutative gauge theory},'' \href{http://arxiv.org/abs/1610.04144}{{\ttfamily arXiv:1610.04144 [hep-th]}}.

\bibitem{Luo:2014sva}
Y.~Luo, M.-C. Tan, J.~Yagi, and Q.~Zhao, ``{Omega-deformation of B-twisted gauge theories and the 3d-3d correspondence},'' \href{http://dx.doi.org/10.1007/JHEP02(2015)047}{{\em JHEP} {\bfseries 02} (2015) 047}, \href{http://arxiv.org/abs/1410.1538}{{\ttfamily arXiv:1410.1538 [hep-th]}}.

\bibitem{Costello:2018txb}
K.~Costello and J.~Yagi, ``{Unification of integrability in supersymmetric gauge theories},'' \href{http://dx.doi.org/10.4310/ATMP.2020.v24.n8.a1}{{\em Adv. Theor. Math. Phys.} {\bfseries 24} no.~8, (2020) 1931--2041}, \href{http://arxiv.org/abs/1810.01970}{{\ttfamily arXiv:1810.01970 [hep-th]}}.

\bibitem{Oh:2021bwi}
J.~Oh and Y.~Zhou, ``{A domain wall in twisted M-theory},'' \href{http://dx.doi.org/10.21468/SciPostPhys.11.4.077}{{\em SciPost Phys.} {\bfseries 11} no.~4, (2021) 077}, \href{http://arxiv.org/abs/2105.09537}{{\ttfamily arXiv:2105.09537 [hep-th]}}.

\bibitem{Witten:1988hf}
E.~Witten, ``{Quantum Field Theory and the Jones Polynomial},'' {\em Commun. Math. Phys.} {\bfseries 121} (1989) 351--399.

\bibitem{Inami:1996zq}
T.~Inami, H.~Kanno, T.~Ueno, and C.-S. Xiong, ``{Two toroidal Lie algebra as current algebra of four-dimensional Kahler WZW model},'' \href{http://dx.doi.org/10.1016/S0370-2693(97)00260-8}{{\em Phys. Lett. B} {\bfseries 399} (1997) 97--104}, \href{http://arxiv.org/abs/hep-th/9610187}{{\ttfamily arXiv:hep-th/9610187}}.

\bibitem{Losev:1995cr}
A.~Losev, G.~W. Moore, N.~Nekrasov, and S.~Shatashvili, ``{Four-dimensional avatars of two-dimensional RCFT},'' \href{http://dx.doi.org/10.1016/0920-5632(96)00015-1}{{\em Nucl. Phys. B Proc. Suppl.} {\bfseries 46} (1996) 130--145}, \href{http://arxiv.org/abs/hep-th/9509151}{{\ttfamily arXiv:hep-th/9509151}}.

\bibitem{Witten:1983tw}
E.~Witten, ``{Global Aspects of Current Algebra},'' \href{http://dx.doi.org/10.1016/0550-3213(83)90063-9}{{\em Nucl. Phys. B} {\bfseries 223} (1983) 422--432}.

\bibitem{Ashwinkumar:2020gxt}
M.~Ashwinkumar, ``{Integrable Lattice Models and Holography},'' \href{http://dx.doi.org/10.1007/JHEP02(2021)227}{{\em JHEP} {\bfseries 02} (2021) 227}, \href{http://arxiv.org/abs/2003.08931}{{\ttfamily arXiv:2003.08931 [hep-th]}}.

\bibitem{Chu:2000bz}
C.-S. Chu, ``{Induced Chern-Simons and WZW action in non-commutative space-time},'' \href{http://dx.doi.org/10.1016/S0550-3213(00)00246-7}{{\em Nucl. Phys. B} {\bfseries 580} (2000) 352--362}, \href{http://arxiv.org/abs/hep-th/0003007}{{\ttfamily arXiv:hep-th/0003007}}.

\bibitem{Gorsky:2001iq}
A.~Gorsky, I.~I. Kogan, and C.~Korthels-Altes, ``{Dualities in quantum Hall system and non-commutative Chern-Simons theory},'' \href{http://dx.doi.org/10.1088/1126-6708/2002/01/002}{{\em JHEP} {\bfseries 01} (2002) 002}, \href{http://arxiv.org/abs/hep-th/0111013}{{\ttfamily arXiv:hep-th/0111013}}.

\bibitem{Cattaneo:2012qu}
A.~S. Cattaneo, P.~Mnev, and N.~Reshetikhin, ``{Classical BV theories on manifolds with boundary},'' \href{http://dx.doi.org/10.1007/s00220-014-2145-3}{{\em Commun. Math. Phys.} {\bfseries 332} (2014) 535--603}, \href{http://arxiv.org/abs/1201.0290}{{\ttfamily arXiv:1201.0290 [math-ph]}}.

\bibitem{Cattaneo:2015vsa}
A.~S. Cattaneo, P.~Mnev, and N.~Reshetikhin, ``{Perturbative quantum gauge theories on manifolds with boundary},'' \href{http://dx.doi.org/10.1007/s00220-017-3031-6}{{\em Commun. Math. Phys.} {\bfseries 357} no.~2, (2018) 631--730}, \href{http://arxiv.org/abs/1507.01221}{{\ttfamily arXiv:1507.01221 [math-ph]}}.

\bibitem{demello}
R.~de~Mello~Koch, K.~Oh, and R.~Tatar, ``{Moduli space for conifolds as intersection of orthogonal D6 branes},'' {\em Nucl. Phys. B} {\bfseries 555} no.~3, (1999) 457--476.

\bibitem{Costello:2016mgj}
K.~Costello and S.~Li, ``{Twisted supergravity and its quantization},'' \href{http://arxiv.org/abs/1606.00365}{{\ttfamily arXiv:1606.00365 [hep-th]}}.

\bibitem{DelZotto:2021ydd}
M.~Del~Zotto, J.~Oh, and Y.~Zhou, ``{Evidence for an algebra of G$_{2}$ instantons},'' \href{http://dx.doi.org/10.1007/JHEP08(2022)214}{{\em JHEP} {\bfseries 08} (2022) 214}, \href{http://arxiv.org/abs/2109.01110}{{\ttfamily arXiv:2109.01110 [hep-th]}}.

\bibitem{Oh:2022unv}
J.~Oh and Y.~Zhou, ``{Evidence for an algebra of $G_2$ instantons II},'' \href{http://arxiv.org/abs/2208.09442}{{\ttfamily arXiv:2208.09442 [hep-th]}}.

\bibitem{witcsg}
E.~Witten, ``{Chern-Simons gauge theory as a string theory},'' {\em Prog. Math.} {\bfseries 133} (1995) 637, \href{http://arxiv.org/abs/hep-th/9207094}{{\ttfamily arXiv:hep-th/9207094}}.

\bibitem{yamazaki}
M.~Yamazaki, ``{New T-duality for Chern-Simons Theory},'' {\em JHEP} {\bfseries 90} (2010) , \href{http://arxiv.org/abs/1904.04976}{{\ttfamily arXiv:1904.04976 [hep-th]}}.

\bibitem{Mikhaylov:2014aoa}
V.~Mikhaylov and E.~Witten, ``{Branes And Supergroups},'' \href{http://dx.doi.org/10.1007/s00220-015-2449-y}{{\em Commun. Math. Phys.} {\bfseries 340} no.~2, (2015) 699--832}, \href{http://arxiv.org/abs/1410.1175}{{\ttfamily arXiv:1410.1175 [hep-th]}}.

\bibitem{Ishtiaque:2021jan}
N.~Ishtiaque, S.~F. Moosavian, S.~Raghavendran, and J.~Yagi, ``{Superspin chains from superstring theory},'' \href{http://dx.doi.org/10.21468/SciPostPhys.13.4.083}{{\em SciPost Phys.} {\bfseries 13} no.~4, (2022) 083}, \href{http://arxiv.org/abs/2110.15112}{{\ttfamily arXiv:2110.15112 [hep-th]}}.

\bibitem{Ashwinkumar:2019mtj}
M.~Ashwinkumar and M.-C. Tan, ``{Unifying lattice models, links and quantum geometric Langlands via branes in string theory},'' \href{http://dx.doi.org/10.4310/ATMP.2020.v24.n7.a1}{{\em Adv. Theor. Math. Phys.} {\bfseries 24} no.~7, (2020) 1681--1721}, \href{http://arxiv.org/abs/1910.01134}{{\ttfamily arXiv:1910.01134 [hep-th]}}.

\bibitem{Ashwinkumar:2018tmm}
M.~Ashwinkumar, M.-C. Tan, and Q.~Zhao, ``{Branes and Categorifying Integrable Lattice Models},'' \href{http://dx.doi.org/10.4310/ATMP.2020.v24.n1.a1}{{\em Adv. Theor. Math. Phys.} {\bfseries 24} no.~1, (2020) 1--24}, \href{http://arxiv.org/abs/1806.02821}{{\ttfamily arXiv:1806.02821 [hep-th]}}.

\bibitem{Aharony:1997bh}
O.~Aharony, A.~Hanany, and B.~Kol, ``{Webs of (p,q) five-branes, five-dimensional field theories and grid diagrams},'' \href{http://dx.doi.org/10.1088/1126-6708/1998/01/002}{{\em JHEP} {\bfseries 01} (1998) 002}, \href{http://arxiv.org/abs/hep-th/9710116}{{\ttfamily arXiv:hep-th/9710116}}.

\bibitem{Leung:1997tw}
N.~C. Leung and C.~Vafa, ``{Branes and toric geometry},'' \href{http://dx.doi.org/10.4310/ATMP.1998.v2.n1.a4}{{\em Adv. Theor. Math. Phys.} {\bfseries 2} (1998) 91--118}, \href{http://arxiv.org/abs/hep-th/9711013}{{\ttfamily arXiv:hep-th/9711013}}.

\bibitem{Li:2020rij}
W.~Li and M.~Yamazaki, ``{Quiver Yangian from Crystal Melting},'' \href{http://dx.doi.org/10.1007/JHEP11(2020)035}{{\em JHEP} {\bfseries 11} (2020) 035}, \href{http://arxiv.org/abs/2003.08909}{{\ttfamily arXiv:2003.08909 [hep-th]}}.

\bibitem{Butson:2023eid}
D.~Butson and M.~Rapcak, ``{Perverse coherent extensions on Calabi-Yau threefolds and representations of cohomological Hall algebras},'' \href{http://arxiv.org/abs/2309.16582}{{\ttfamily arXiv:2309.16582 [math.RT]}}.

\bibitem{Aganagic}
M.~Aganagic, K.~Costello, J.~McNamara, and C.~Vafa, ``{Topological Chern-Simons Matter Theories},'' \href{http://arxiv.org/abs/1706.09977}{{\ttfamily arXiv:1706.09977 [hep-th]}}.

\bibitem{Moore:1989yh}
G.~W. Moore and N.~Seiberg, ``{Taming the Conformal Zoo},'' {\em Phys. Lett. B} {\bfseries 220} (1989) 422--430.

\bibitem{Elitzur:1989nr}
S.~Elitzur, G.~W. Moore, A.~Schwimmer, and N.~Seiberg, ``{Remarks on the Canonical Quantization of the Chern-Simons-Witten Theory},'' \href{http://dx.doi.org/10.1016/0550-3213(89)90436-7}{{\em Nucl. Phys. B} {\bfseries 326} (1989) 108--134}.

\bibitem{Chu:2009ms}
C.-S. Chu and D.~J. Smith, ``{Multiple Self-Dual Strings on M5-Branes},'' \href{http://dx.doi.org/10.1007/JHEP01(2010)001}{{\em JHEP} {\bfseries 01} (2010) 001}, \href{http://arxiv.org/abs/0909.2333}{{\ttfamily arXiv:0909.2333 [hep-th]}}.

\bibitem{Berman:2009xd}
D.~S. Berman, M.~J. Perry, E.~Sezgin, and D.~C. Thompson, ``{Boundary Conditions for Interacting Membranes},'' \href{http://dx.doi.org/10.1007/JHEP04(2010)025}{{\em JHEP} {\bfseries 04} (2010) 025}, \href{http://arxiv.org/abs/0912.3504}{{\ttfamily arXiv:0912.3504 [hep-th]}}.

\bibitem{Aharony:2008ug}
O.~Aharony, O.~Bergman, D.~L. Jafferis, and J.~Maldacena, ``{N=6 superconformal Chern-Simons-matter theories, M2-branes and their gravity duals},'' \href{http://dx.doi.org/10.1088/1126-6708/2008/10/091}{{\em JHEP} {\bfseries 10} (2008) 091}, \href{http://arxiv.org/abs/0806.1218}{{\ttfamily arXiv:0806.1218 [hep-th]}}.

\bibitem{Faizal:2016skd}
M.~Faizal, Y.~Luo, D.~J. Smith, M.-C. Tan, and Q.~Zhao, ``{Gauge and supersymmetry invariance of $\mathcal{N}=2$ boundary Chern–Simons theory},'' \href{http://dx.doi.org/10.1016/j.nuclphysb.2016.11.020}{{\em Nucl. Phys. B} {\bfseries 914} (2017) 577--598}, \href{http://arxiv.org/abs/1601.05429}{{\ttfamily arXiv:1601.05429 [hep-th]}}.

\bibitem{Faizal:2011cd}
M.~Faizal and D.~J. Smith, ``{Supersymmetric Chern-Simons Theory in Presence of a Boundary},'' \href{http://dx.doi.org/10.1103/PhysRevD.85.105007}{{\em Phys. Rev. D} {\bfseries 85} (2012) 105007}, \href{http://arxiv.org/abs/1112.6070}{{\ttfamily arXiv:1112.6070 [hep-th]}}.

\bibitem{Berman:2009kj}
D.~S. Berman and D.~C. Thompson, ``{Membranes with a boundary},'' \href{http://dx.doi.org/10.1016/j.nuclphysb.2009.06.004}{{\em Nucl. Phys. B} {\bfseries 820} (2009) 503--533}, \href{http://arxiv.org/abs/0904.0241}{{\ttfamily arXiv:0904.0241 [hep-th]}}.

\bibitem{Cattaneo:2020lle}
A.~S. Cattaneo, P.~Mnev, and K.~Wernli, ``{Quantum Chern–Simons Theories on Cylinders: BV-BFV Partition Functions},'' \href{http://dx.doi.org/10.1007/s00220-022-04513-8}{{\em Commun. Math. Phys.} {\bfseries 398} (2023) 133--218}, \href{http://arxiv.org/abs/2012.13983}{{\ttfamily arXiv:2012.13983 [hep-th]}}.

\bibitem{Okazaki:2015fiq}
T.~Okazaki and D.~J. Smith, ``{Topological M-strings and supergroup Wess-Zumino-Witten models},'' \href{http://dx.doi.org/10.1103/PhysRevD.94.065016}{{\em Phys. Rev. D} {\bfseries 94} no.~6, (2016) 065016}, \href{http://arxiv.org/abs/1512.06646}{{\ttfamily arXiv:1512.06646 [hep-th]}}.

\bibitem{Susskind:2001fb}
L.~Susskind, ``{The Quantum Hall fluid and non-commutative Chern-Simons theory},'' \href{http://arxiv.org/abs/hep-th/0101029}{{\ttfamily arXiv:hep-th/0101029}}.

\bibitem{Hellerman:2001rj}
S.~Hellerman and M.~Van~Raamsdonk, ``{Quantum Hall physics equals non-commutative field theory},'' \href{http://dx.doi.org/10.1088/1126-6708/2001/10/039}{{\em JHEP} {\bfseries 10} (2001) 039}, \href{http://arxiv.org/abs/hep-th/0103179}{{\ttfamily arXiv:hep-th/0103179}}.

\bibitem{Daoud:2006nf}
M.~Daoud and A.~Jellal, ``{Effective Weiss-Zumino-Witten Action for Edge States of Quantum Hall Systems on Bargman Ball},'' \href{http://dx.doi.org/10.1016/j.nuclphysb.2006.11.032}{{\em Nucl. Phys. B} {\bfseries 764} (2007) 109--127}, \href{http://arxiv.org/abs/hep-th/0605289}{{\ttfamily arXiv:hep-th/0605289}}.

\bibitem{Karabali:2004km}
D.~Karabali and V.~Nair, ``{Edge states for quantum Hall droplets in higher dimensions and a generalized WZW model},'' \href{http://dx.doi.org/10.1016/j.nuclphysb.2004.07.014}{{\em Nucl. Phys. B} {\bfseries 697} (2004) 513--540}, \href{http://arxiv.org/abs/hep-th/0403111}{{\ttfamily arXiv:hep-th/0403111}}.

\bibitem{Hu:2023eyx}
S.~Hu, S.~Li, D.~Ye, and Y.~Zhou, ``{Quantum Algebra of Chern-Simons Matrix Model and Large $N$ Limit},'' \href{http://arxiv.org/abs/2308.14046}{{\ttfamily arXiv:2308.14046 [math.QA]}}.

\bibitem{Hu:2024waw}
S.~Hu, S.~Li, D.~Ye, and Y.~Zhou, ``{On the Hilbert Space of the Chern-Simons Matrix Model, Deformed Double Current Algebra Action, and the Conformal Limit},'' \href{http://arxiv.org/abs/2409.12486}{{\ttfamily arXiv:2409.12486 [math-ph]}}.

\bibitem{Haouzi:2024qyo}
N.~Haouzi and S.~Jeong, ``{Miura operators as R-matrices from M-brane intersections},'' \href{http://arxiv.org/abs/2407.15990}{{\ttfamily arXiv:2407.15990 [hep-th]}}.

\bibitem{Ashwinkumar:2024vys}
M.~Ashwinkumar, ``{R-matrices from Feynman Diagrams in 5d Chern-Simons Theory and Twisted M-theory},'' \href{http://arxiv.org/abs/2408.15732}{{\ttfamily arXiv:2408.15732 [hep-th]}}.

\bibitem{Ishtiaque:2024orn}
N.~Ishtiaque, S.~Jeong, and Y.~Zhou, ``{R-matrices and Miura operators in 5d Chern-Simons theory},'' \href{http://arxiv.org/abs/2408.15712}{{\ttfamily arXiv:2408.15712 [hep-th]}}.

\bibitem{Donaldson:1985zz}
S.~K. Donaldson, ``{ANTI SELF-DUAL YANG-MILLS CONNECTIONS OVER COMPLEX ALGEBRAIC SURFACES AND STABLE VECTOR BUNDLES},'' \href{http://dx.doi.org/10.1112/plms/s3-50.1.1}{{\em Proc. Lond. Math. Soc.} {\bfseries 50} (1985) 1--26}.

\bibitem{Nair:1990aa}
V.~P. Nair and J.~Schiff, ``{A Kahler-{Chern-Simons} Theory and Quantization of Instanton Moduli Spaces},'' \href{http://dx.doi.org/10.1016/0370-2693(90)90624-F}{{\em Phys. Lett. B} {\bfseries 246} (1990) 423--429}.

\bibitem{Costello:2023hmi}
K.~Costello, N.~M. Paquette, and A.~Sharma, ``{Burns space and holography},'' \href{http://dx.doi.org/10.1007/JHEP10(2023)174}{{\em JHEP} {\bfseries 10} (2023) 174}, \href{http://arxiv.org/abs/2306.00940}{{\ttfamily arXiv:2306.00940 [hep-th]}}.

\end{thebibliography}\endgroup

\end{document}